\documentclass[conference]{IEEEtran}
\IEEEoverridecommandlockouts

\usepackage{cite}
\usepackage{amsmath,amssymb,amsfonts}
\usepackage{algorithmic}
\usepackage{graphicx}
\usepackage{textcomp}
\usepackage{xcolor}
\usepackage[ruled,vlined,linesnumbered]{algorithm2e}
\usepackage{caption,subcaption}
\usepackage{multirow}
\usepackage{adjustbox}
\usepackage{pifont}
\usepackage{url}
\usepackage{soul}
\usepackage{threeparttable}
\usepackage{float}
\usepackage{booktabs}
\usepackage[utf8]{inputenc}
\usepackage[T1]{fontenc}
\usepackage[english]{babel}
\usepackage{mwe}
\usepackage{capt-of}
\usepackage{colortbl}
\usepackage{setspace}
\usepackage{tcolorbox}
\usepackage{hyperref}
\usepackage{balance}

\usepackage{tikz}
\newcommand*\circled[1]{\tikz[baseline=(char.base)]{
            \node[shape=circle,draw,inner sep=2pt] (char) {#1};}}

\def\BibTeX{{\rm B\kern-.05em{\sc i\kern-.025em b}\kern-.08em
    T\kern-.1667em\lower.7ex\hbox{E}\kern-.125emX}}

\hypersetup{
colorlinks=true,
linkcolor=black,
citecolor=black
}
    
\newcommand{\fullset}{FullSet}
\newcommand{\randomset}{RandomSet}

\newcommand{\covbase}{CISS}
\newcommand{\minset}{\covbase{}$_{M}$}
\newcommand{\peachset}{\covbase{}$_{P}$}

\newcommand{\fuzzbase}{PISS}
\newcommand{\hotset}{\fuzzbase{}}

\newcommand{\feabase}{FISS}
\newcommand{\tfidfset}{\feabase{}$_{\textit{TS}}$}
\newcommand{\astset}{\feabase{}$_{\textit{AST}}$}
\newcommand{\cfgset}{\feabase{}$_{\textit{CFG}}$}
\newcommand{\codetset}{\feabase{}$_{\textit{CT}}$}
\newcommand{\plbartset}{\feabase{}$_{\textit{PB}}$}
\newcommand{\infercodeset}{\feabase{}$_{\textit{IC}}$}
\newcommand{\codebertset}{\feabase{}$_{\textit{CB}}$}

\newcommand{\unknownBug}{25}
\newcommand{\confirmBug}{21}
\newcommand{\newCorpusBug}{7}

\newcommand{\Comment}[1]{}

\newcommand\addCitation[1]{\textit{\textcolor{red}{[citations]}}}

\begin{document}

\title{Selecting Initial Seeds for Better JVM Fuzzing
\thanks{\IEEEauthorrefmark{1} Junjie Chen is the corresponding author.}
}

\author{

\IEEEauthorblockN{Tianchang Gao}
\IEEEauthorblockA{
\textit{College of Intelligence and Computing} \\
\textit{Tianjin University}\\
Tianjin, China \\
gaotc090@tju.edu.cn
}

\and

\IEEEauthorblockN{Junjie Chen\IEEEauthorrefmark{1}}
\IEEEauthorblockA{\textit{College of Intelligence and Computing} \\
\textit{Tianjin University}\\
Tianjin, China \\
junjiechen@tju.edu.cn}

\and

\IEEEauthorblockN{Dong Wang}
\IEEEauthorblockA{\textit{College of Intelligence and Computing} \\
\textit{Tianjin University}\\
Tianjin, China \\
dong\_w@tju.edu.cn}

\and

\IEEEauthorblockN{Yile Guo}
\IEEEauthorblockA{\textit{College of Intelligence and Computing} \\
\textit{Tianjin University}\\
Tianjin, China \\
gyl\_666@tju.edu.cn}

\and

\IEEEauthorblockN{Yingquan Zhao}
\IEEEauthorblockA{\textit{College of Intelligence and Computing} \\
\textit{Tianjin University}\\
Tianjin, China \\
zhaoyingquan@tju.edu.cn}

\and

\IEEEauthorblockN{Zan Wang}
\IEEEauthorblockA{\textit{College of Intelligence and Computing} \\
\textit{Tianjin University}\\
Tianjin, China \\
wangzan@tju.edu.cn}
}

\maketitle

\begin{abstract}
JVM fuzzing techniques serve as a cornerstone for guaranteeing the quality of implementations. 
In typical fuzzing workflows, initial seeds are crucial as they form the basis of the process. 
Literature in traditional program fuzzing has confirmed that effectiveness is largely impacted by redundancy among initial seeds, thereby proposing a series of seed selection methods.
JVM fuzzing, compared to traditional ones, presents unique characteristics, including large-scale and intricate code, and programs with both syntactic and semantic features.
However, it remains unclear whether the existing initial seed selection methods are suitable for JVM fuzzing and whether utilizing program features can enhance effectiveness.
To address this, we devise a total of 10 initial seed selection methods, comprising coverage-based, prefuzz-based, and program-feature-based methods.
We then conduct an empirical study on three JVM implementations to extensively evaluate the performance of the initial seed selection methods within two state-of-the-art fuzzing techniques (JavaTailor and VECT).
Specifically, we examine performance from three aspects: (i) effectiveness and efficiency using widely studied initial seeds, (ii) effectiveness using the programs in the wild, and (iii) the ability to detect new bugs.
Evaluation results first show that the program-feature-based method that utilizes the control flow graph not only has a significantly lower time overhead (i.e., 30s), but also outperforms other methods, achieving 142\% to 269\% improvement compared to the full set of initial seeds.
Second, results reveal that the initial seed selection greatly improves the quality of wild programs and exhibits complementary effectiveness by detecting new behaviors.
Third, results demonstrate that given the same testing period, initial seed selection improves the JVM fuzzing techniques by detecting more unknown bugs. 
Particularly, \confirmBug{} out of the 25 detected bugs have been confirmed or fixed by developers.
This work takes the first look at initial seed selection in JVM fuzzing, confirming its importance in fuzzing effectiveness and efficiency. 
\end{abstract}

\begin{IEEEkeywords}
Java Virtual Machine, JVM Fuzzing, Initial Seed Selection, Empirical Study
\end{IEEEkeywords}

\section{Introduction}

The Java Virtual Machine (JVM) serves as a critical infrastructure for the Java platform, providing a consistent runtime environment that allows developers to create and deploy Java applications across diverse hardware and operating systems without needing modification. 
Given its foundational role, numerous JVM fuzzing techniques have been developed to ensure the quality of JVM implementations~\cite{classming,javatailor,vect}. 
These techniques typically follow a general workflow, which involves (1) preparing a set of initial seeds, (2) generating test programs through mutations or code synthesis applied to these seeds, and (3) testing JVMs with the generated test programs and providing feedback for subsequent iterations.
Among these steps, the initial seeds form the basis of the entire fuzzing process and thus play a critical role in determining the overall effectiveness of the fuzzing campaign.

In practice, fuzzing budgets are often constrained, and redundancy among initial seeds, concerning the behaviors they induce during the fuzzing process, is a common occurrence~\cite{2014selection}. 
This redundancy necessitates spending a significant amount of time on initial seeds with overlapping behaviors for fuzzing, consequently limiting overall effectiveness within the allocated time budget~\cite{2021selection}.
Several studies have been conducted to confirm such an influence on fuzzing \textit{traditional programs}, and proposed some methods of selecting a subset of initial seeds for improving the fuzzing effectiveness, such as coverage-based~\cite{peachfuzzer,2021selection} and prefuzz-based~\cite{2014selection} methods.

JVM fuzzing presents unique characteristics distinct from traditional program fuzzing. 
On the one hand, JVM code tends to be large-scale and intricate~\cite{chen2020survey}, resulting in significant overhead in collecting coverage and executing test programs. 
Therefore, both coverage-based and prefuzz-based methods suffer from cost issues.
On the other hand, JVM fuzzing operates on programs containing rich syntactic and semantic features, and several existing works have highlighted the correlation between test program features and JVM fuzzing effectiveness~\cite{vect,jopfuzzer}. 
Given these unique characteristics, \textit{it remains uncertain whether existing initial seed selection methods are suitable for JVM fuzzing and whether incorporating test program features can enhance initial seed selection effectiveness}. 
To address these open questions, we conducted an empirical study to investigate the influence of initial seed selection in the context of JVM fuzzing, with the aim of further improving the effectiveness of JVM fuzzing within limited time budgets.

In this study, we investigated a total of 10 initial seed selection methods, which comprised two coverage-based methods (\textbf{\covbase{}})~\cite{peachfuzzer,2014selection}, one prefuzz-based method (\textbf{\fuzzbase{}})~\cite{2014selection}, and seven program-feature-based methods (\textbf{\feabase{}}).
% In this study, we investigated a total of 10 initial seed selection methods, which comprised two \textbf{C}overage-based~\cite{peachfuzzer,2014selection}, one \textbf{P}refuzz-based~\cite{2014selection} and seven Program-\textbf{F}eature-based \textbf{I}nitial \textbf{S}eed \textbf{S}election methods, i.e., \textbf{\covbase{}}, \textbf{\fuzzbase{}} and \textbf{\feabase{}}, respectively.
The coverage-based and prefuzz-based methods were originally proposed in the context of traditional program fuzzing, and we adopted them for JVM fuzzing.
Specifically, they utilize test coverage and short-time fuzzing results as metrics to assess the fuzzing capability of each initial seed for selection, respectively.
The remaining seven methods were particularly designed by us for JVM fuzzing, considering different ways of utilizing program features.
Specifically, these methods utilize textual features, Abstract Syntax Tree (AST) features, Control Flow Graph (CFG) features, and code semantics obtained from various code representation models to measure the diversity of initial seeds for selection, respectively.
Further details of these studied methods will be presented in Section~\ref{sec:corpusreduction}.

Based on the set of methods, we performed an experiment to address the following three research questions.\\
\noindent
(\textbf{RQ1}): \ul{How do these initial seed selection methods perform in JVM fuzzing?}
We first seek to investigate the performance of the studied initial seed selection methods in terms of effectiveness and efficiency.
In particular, we used three widely-used JVM implementations (i.e., HotSpot~\cite{hotspot}, OpenJ9~\cite{openj9}, and Bisheng JDK~\cite{bisheng}) as subjects, two widely-used sets of initial seeds as the ones for selection, and two state-of-the-art JVM fuzzing techniques (i.e., JavaTailor~\cite{javatailor} and VECT~\cite{vect}) for evaluating each selected subset of initial seeds. 
To make our conclusions have statistical significance, we used the historical versions of these JVM implementations as the studied JVM fuzzing techniques can detect more bugs in them.\\
\noindent
\textbf{Results:}
% The program-feature-based method that utilizes the control flow graph, \cfgset{}, not only enhances the effectiveness of JVM fuzzing by detecting more inconsistencies
% with a relatively small corpus subset but also has a significantly lower time overhead compared to other methods.
% Specifically, we recommend setting the corpus budget for \cfgset{} between 35\% and 50\% to achieve optimal performance and resource utilization in JVM fuzzing.
% \jj{the term of inconsistency cannot be used without definition} 
% \jj{some quantative results are required.}
% \jj{we should list major results and conclusions. this conclusion is a bit weak.}
The program-feature-based method that utilizes CFG features, \cfgset{}, not only has a significantly lower time overhead (i.e., 30s), but also outperforms other methods, achieving 142\% to 269\% improvement compared to the full set of initial seeds.
Specifically, we recommend using \cfgset{} with a 50\% budget for initial seed selection, which can optimize performance and resource utilization in JVM fuzzing.
\\
\noindent
(\textbf{RQ2}): \ul{How does initial seed selection perform on programs in the wild for improving JVM fuzzing?}
% \zyq{"program in the wild" sounds wired, maybe we can use "real-world programs" or "programs in practice" ?} \Wang{wild should be fine, a buzz word recently.}
Intuitively, any Java programs in the wild can be used as initial seeds for JVM fuzzing.
It is plausible that some of these programs may introduce new behaviors compared to widely-studied initial seeds, thereby potentially improving JVM fuzzing. 
However, there is also a substantial number of programs that may not introduce new aspects for JVM fuzzing. 
Randomly collecting a large number of programs in the wild as initial seeds is unlikely to significantly enhance the effectiveness of JVM fuzzing within limited time budgets. 
This may be a primary reason why existing JVM fuzzing techniques do not use them as initial seeds.
However, the existence of initial seed selection methods presents an opportunity to strategically incorporate programs in the wild. 
This may help leverage the benefits of such programs effectively and thus improve JVM testing. 
Therefore, we designed RQ2 in our study, which not only investigates the generalizability of our conclusions but also explores the feasibility of a new aspect to enhance the effectiveness of JVM fuzzing. 
Specifically, we collected a set of programs in the wild as a new corpus and then repeated the first experiment on it.\\
\noindent
\textbf{Results:}
% \jj{Main findings can be reported here.}
% Programs in the wild complement the widely-studied initial seeds by detecting certain inconsistencies.
% In addition, \cfgset{} remains the most effective seed selection method. 
% These results not only prove the significance of expanding the corpus but also highlight the generalizability of \cfgset{}.
Programs in the wild complement the widely-studied initial seeds by detecting new JVM behaviors.
The subset of wild programs selected by \cfgset{} boosts detection capability, achieving an improvement of 8.3\% to 108\% compared to the full set of initial seeds.
Moreover, results underscore the crucial role of initial seed selection in enhancing the effectiveness of wild programs, and also reinforce the generalizability of the top-performing method, \cfgset{}.
\\
\noindent
(\textbf{RQ3}): \ul{Can initial seed selection survive existing JVM fuzzing techniques for detecting new bugs?}
% \jj{We have a core topic: initial seed selection. Thus, we should highlight it in each RQ, rather than make it equal to open-source corpus.}
Based on RQ1 and RQ2 results, it is evident that initial seed selection is helpful in enhancing the effectiveness of JVM fuzzing by allocating more time to diverse (selected) seeds and enabling the effective incorporation of programs in the wild. However, the ultimate aim of fuzzing is to find bugs in JVM, thus
we further investigated whether initial seed selection also helps improve studied fuzzing techniques in detecting previously unknown bugs.
Specifically, we applied each subset of initial seeds selected by different seed selection methods on VECT to test the latest versions of three JVM implementations. We then compared the number of unknown bugs detected by each method (including the full set) within the same testing period.
% \Wang{``applied n. (subset) to n. (JVM implementations)'' it is not clear to me. ``applied state-of-the-art fuzzer on the latest versions of three JVM implementations by selecting ...''?}
\\
\noindent
\textbf{Results:}
% \jj{Main findings can be reported here.}
% The use of initial seed selection and open-source corpus improves JVM techniques does enhance JVM techniques for detecting previously unknown bugs. In particular, we identified a total of \unknownBug{} bugs, out of which \confirmBug{} bugs have been confirmed or fixed by developers.
% \jj{I cannot understand this conclusion according to the experimental setup. I feel they don't match with each other.}
Initial seed selection \cfgset{} improves the JVM fuzzing techniques by detecting more previously unknown bugs within the same testing period.
Of the \unknownBug{} submitted bugs, the developers have confirmed and fixed \confirmBug{} bugs, and \newCorpusBug{} can only be detected using initial seeds from wild programs.
% Within the same testing period, \cfgset{} not only detected all the bugs found by the full set of initial seeds but also detected an additional 6 previously unknown bugs.

% \jj{List our main implications from our study.}

To sum up, our major contributions are four-fold:
\begin{itemize}
    \item Design of seven initial seed selection methods specifically tailored for JVM fuzzing, leveraging various features in test programs.

    \item Implementation of all studied initial seed selection methods as an open-source toolkit\cite{homepage}, facilitating practical use and future research in the field.

    \item Conducting an extensive study to investigate the impact of initial seed selection on the effectiveness of JVM fuzzing, addressing three research questions (RQs). 
    % \Wang{could incorporate the contribution of the open-source corpus.}
    % \jj{I think we should emphasize one topic. the effectiveness of open-source corpus is brought by initial seed selection in actual.}

    \item Obtaining a series of findings and implications for improving JVM fuzzing. 
    % In particular, we detected \unknownBug{} previously unknown bugs, \confirmBug{} of which have been confirmed or fixed by developers. 
    % \Wang{The description of bugs is repetitive. Should be more high-level.}
\end{itemize}

\section{Background and Related Work}
\label{sec:background}
% In this section, we introduce the background and related work regarding JVM fuzzing and initial seed selection. 

\subsection{JVM Fuzzing}
% \jj{I don't suggest to highlight this classification since both our studied techniques belong to the same category. We can introduce the two studied techniques in detail, and then briefly introduce some other techniques that also require initial seeds. We then highlight that our work is general to various JVM fuzzing techniques that require initial seeds. Finally, we can mention that there are some techniques without the requirement on initial seeds.}

Fuzzing is one of the most popular and effective methods to find bugs and vulnerabilities in  software~\cite{chen2020survey,fuzzing1,fuzzing2,fuzzing3}. 
In the context of JVM, diverse fuzzing techniques have been proposed to ensure the quality of JVM implementations.
The majority of them are seed-driven, such as the state-of-the-art \textit{JavaTailor} and \textit{VECT} that are experimented in this study.\\
\noindent
- \textbf{JavaTailor} adopts a history-driven approach for synthesizing test programs. 
It designs five types of ingredients, extracted from bug-revealing test programs. 
These ingredients are then inserted into seed programs, and syntactic and semantic constraints within the constituents are automatically rectified to produce valid synthesized programs for differential testing.\\
\noindent
-  \textbf{VECT} streamlines the extensive ingredient space by vectorizing program ingredients through advanced code representation. It then employs a feedback-based ingredient selection strategy to enhance the effectiveness of test program generation. 
Furthermore, VECT incorporates an enhanced test oracle to broaden the bug detection capability of current JVM testing. 

Additionally, there are other techniques that are also seed-driven.
For example, Chen et al. introduced \textit{classfuzz}~\cite{classfuzz} and \textit{classming}~\cite{classming}, which mutate bytecode and control flow to generate test programs with broader coverage, respectively.
Wu et al. proposed \textit{JITfuzz}~\cite{jitfuzz}, which designs mutation operators related to JIT defects to generate test programs.
\textit{SJFuzz}~\cite{sjfuzz} optimizes the scheduling of seed programs and mutation operators to enhance the effectiveness of classming. 

Apart from the seed-driven techniques, some efforts target generating test programs by starting from an empty program or a program with holes~\cite{javafuzzer,justgen,jattack}.
For instance, Zhang et al. proposed \textit{JAttack}~\cite{jattack}, which fills holes in template classes with randomly generated expressions and values to generate test programs. 
Hwang et al. introduced \textit{JUSTGen}~\cite{justgen}, which uses JNI specifications to identify unspecified scenarios for generating corner cases.

Our work is applicable to various JVM fuzzing techniques that require initial seeds.
We devised a set of methods for selecting initial seeds, in order to extensively study their effect on fuzzing capability.

\subsection{Initial Seed Selection}
In seed-driven fuzzing workflows, the selection of initial seeds is typically one of the first steps and plays a significant role in the overall effectiveness of fuzzing.
Several studies have highlighted the importance of high-quality seeds on fuzzers' performance.
For instance, Klees et al. ~\cite{klees2018evaluating} asserted that fuzzer's performance can greatly vary on the same program, depending on the seed used.
Herrera et al. ~\cite{2021selection} evaluated how seed selection affects a fuzzer's ability and their findings suggest that seed selection is a critical step that must be considered prior to launching any fuzzing campaign.
Remarkably, practitioners (such as the developers of the Mozilla Firefox browser~\cite{firefoxfuzzing}) also recognized the importance of seed selection.

Several methods have been proposed to select a subset of initial seeds, with the aim of improving the quality of initial seeds and reducing the budget of fuzzing.
Specifically, a common method in traditional software testing is the corpus minimization technique, which selects the smallest subset of the corpus that maintains the equivalent coverage to the entire corpus.
This technique involves widely-adopted coverage-based methods, such as \textit{PeachSet}~\cite{peachfuzzer}, \textit{MinSet}~\cite{2014selection}, and \textit{OptiMin}~\cite{2021selection}.
For example, MinSet, an unweighted greedy-reduced minimization, sorts seed files based on their coverage increments and chooses the one with the maximum increase.
The prior work reported that PeachSet found the highest number of bugs and MinSet performed the best in terms of the minimization ability~\cite{2014selection}.
The prefuzz-based method \textit{HotSet}~\cite{2014selection} executes each seed file for an equal amount of time and sorts them based on the number of known defects each file uncovers.

% In addition to selecting a subset of the corpus, there is also some works to select and assess other aspects of the seed~\cite{ss4os,ss4js,ss4dnn,targetfuzz}. For example, Pailoor et al. introduced MoonShine~\cite{ss4os}, which utilizes lightweight static analysis to efficiently detect dependencies across different system calls and select the system call that has dependencies in the seed. Additionally, Wen et al.~\cite{ss4js} conducted extensive experiments to systematically evaluate the influence of the seed source on fuzzing JavaScript engines.

Despite extensive study of initial seed selection in traditional software, its role in JVM fuzzing remains largely unexplored. 
To address this, we conducted the first study to investigate the effect of various selection methods on JVM fuzzing performance.
Different from those methods aiming to minimize the initial seeds, as the initial step, our focus is to understand the effect of selecting subsets of the seeds.
Furthermore, certain existing methods (such as HotSet) are unable to achieve minimization. 
Hence, to make a fair comparison among these methods, we opted to establish a selected subset of initial seeds.
Current selection methods may cause considerable overhead.
The code space of JVM is vast, and collecting coverage for each seed program takes about 500 seconds. The time required to apply existing coverage-based methods to a large corpus is substantial. Similarly, techniques based on known bugs require individual fuzzing for each seed program, which further consumes significant time resources.
To mitigate this challenge, we novelly designed several program-feature-based methods and evaluated their performance.

\section{Studied Initial Seed Selection Methods}
\label{sec:corpusreduction}
% \jj{We need to define the selection problem before introducing various selection techniques, which can be presented in Section II.With this definition, the presentation of these methods can be more convenient.}
Drawing inspiration from existing works in the realm of traditional software, we devised a set of initial seed selection methods tailored for JVM testing.
Table~\ref{tab:techniques} presents a summary of the studied ten methods including their types, names, inputs, and search strategies.
These methods are categorized into three groups: coverage-based, prefuzz-based, and program-feature-based methods.
For simplicity, we will use the term corpus to refer to the entire set of initial seeds in the following sections.
% We now introduce the design and implementation of them below. 
% \begin{itemize}
%     \item Section~\ref{subsec:CovTech} presents two coverage-based corpus reduction techniques (\covbase{}), adapted from PeachSet and MinSet. Since OptiMin and EDTSO are used to solve MaxSAT and ILP problems, respectively, and only the solver can be used for corpus minimization, they cannot be used. 

%     \item Section~\ref{subsec:BugTech} introduces a known inconsistency-based corpus reduction technique (\fuzzbase{}), adapted from HotSet. Because fewer crashes are detected In JVM testing, in this study, we use inconsistency as metric for \fuzzbase{}.

%     \item Section~\ref{subsec:FeaTech} proposes program feature-based corpus reduction techniques (\feabase{}), inspired by relevant work in the Test Case Prioritization (TCP) field, and delineates specific feature extraction schemes.
% \end{itemize}

\begin{table}[t]
\small
\centering
\caption{Studied initial seed selection methods}
\label{tab:techniques}
\begin{spacing}{1.2}
\resizebox{.8\linewidth}{!}{
\begin{tabular}{llll}
\toprule
\textbf{Type}                  & \textbf{Method}       & \textbf{Input}             & \textbf{Strategy}         \\ \midrule
\multirow{2}{*}{White-box} & \peachset{}       & Coverage          & Greedy           \\
                           & \minset{}     & Coverage          & Greedy           \\
                           \midrule
                           
\multirow{8}{*}{Black-box} & \hotset{}       & Fuzzing Result   & Greedy           \\
                            \cmidrule(l){2-4}

                           & \tfidfset{}     & Token             & FPS              \\
                           & \astset{}       & AST               & FPS              \\
                           & \cfgset{}       & CFG               & FPS              \\ \cmidrule(l){2-4}
                           
                           & \codebertset{}  & Token             & FPS              \\ 
                           & \infercodeset{} & AST               & FPS              \\
                           & \plbartset{}    & Token             & FPS              \\
                           & \codetset{}     & Token             & FPS              \\ \bottomrule
                           
\end{tabular}
}
\end{spacing}
\vspace{-.4cm}
\end{table}

\subsection{Coverage-based Method}
\label{subsec:CovTech}

Two white-box initial seed selection methods based on coverage metrics are introduced: \textit{\peachset{}} and \textit{\minset{}}.
Like PeachSet and MinSet, \peachset{} sorts seed programs based on coverage metrics, then selects the top-k percent as the corpus subset. On the other hand, \minset{} sorts these programs based on coverage increments and stops the selection process once all candidate seed programs show zero coverage increments.
To accommodate our experiment implementation, when the selection process stops, \minset{} transitions into a mode the same as \peachset{}. In this mode, candidate seed programs are sorted based on coverage metrics. Likewise, the top-k percent of these programs is selected as the corpus subset. 
Note that when several seed programs have the same coverage, we randomly select one of them to break the tie following the existing work~\cite{tie1,2014selection}.

\subsection{Prefuzz-based Method}
\label{subsec:BugTech}
% \jj{Please don't specially highlight the difference between bugs and inconsistencies, which may bring confusion to readers. Also, we still don't define inconsistencies.we shouldn't mix many meanings in a paragraph. we can just introduce the selection method here.}
\textit{\hotset{}} is a black-box method adapted from HotSet. Specifically, \hotset{} performs fuzzing on each seed program for t seconds, recording the number of bugs each seed program uncovers. It then selects the top-k percent with the highest number of bugs to form the corpus subset. Following the existing work~\cite{2014selection}, we conduct fuzzing on each seed program for 5 minutes (t = 300) to compute the subset of initial seeds selected by \hotset{} in our experiments.

\subsection{Program-Feature-based Method}
\label{subsec:FeaTech}
% \jj{We shouldn't compare these methods here. We just introduce why we design program-feature-based methods and their details. We can refer to the reason mentioned in Intro.}
Previous work has emphasized the correlation between test program features and JVM fuzzing effectiveness~\cite{chenicse17,chenicst16}. 
Specifically, similar program features may possess the ability to trigger similar bugs or the absence of bugs. 
Moreover, the triggering of bugs is frequently associated with specific program features rather than common ones.
Hence, we further propose a program-feature-based initial seed selection method (\textit{\feabase{}}) by extracting program features for evaluation. 
% This draws inspiration from the current test case prioritization works~\cite{chenicse17,chenicst16}, based on two concepts: 1) similar program features may represent the ability to trigger similar bugs or the absence of bugs, and 2) the triggering of bugs is frequently associated with specific program features rather than common ones.
Based on whether we extract program features using statistical models or pre-trained models, we categorize them into traditional features and code semantics.
Below we describe the process for extracting program features including traditional features and code semantics, as well as the corpus subset selection.
% \jj{Why we have this classification? We need some reasonable explanation.}

% To address the efficiency limitation of \covbase{} and \fuzzbase{} in JVM testing, inspired by FAST and ATM, we posit two key insights: 1) similar program features may represent the ability to trigger similar bugs or the absence of bugs, and 2) the triggering of bugs is frequently associated with specific program features rather than common ones. In this study, we implemented \feabase{} and extracted seven program features for evaluation.

\textbf{Traditional feature extraction.} 
% \jj{the outputs of this step should be feature vectors, but the current version doesn't mention it explicitly.} 
Three commonly utilized types of information in program analysis are selected: textual information, syntactic information, and semantic information~\cite{tokeninfo,astinfo,cfginfo}.
Specifically, we analyze the Token Sequence (TS), Abstract Syntax Tree (AST), and Control Flow Graph (CFG) of programs to extract traditional features, respectively. 
% \jj{we should clearly separate the presentation of method insight and specific implementation.}
\begin{itemize}
    \item \textit{TS features}: Each token in the corpus is treated as a dimension in the feature vector, and the TF-IDF~\cite{tfidf} is used to compute the score of each token in the seed program. Tokens that are absent receive a value of 0.
    \item \textit{AST features}: The Java syntax analysis tool JDT~\cite{jdt} is utilized to obtain the AST of each seed program and extract tree-based n-gram chains. 
    Existing work~\cite{3gramset} has shown that setting n to 3 yields better representation performance, so we also extract 3-gram chains for our method.
    % \zyq{why 3-gram? should we discuss the setting of this parameter?}
    Each 3-gram chain in the corpus is regarded as a dimension in the feature vector, and the frequency of occurrence of each 3-gram chain in the seed program is counted. Chains that do not exist receive a value of 0.
    \item \textit{CFG features}: The Java bytecode analysis tool Soot~\cite{soot} is used to obtain the CFG and extract graph-based 3-gram chains. 
    Each node in the CFG represents a Jimple instruction since Soot analyzes bytecode using the intermediate language, Jimple. 
    Similar to AST features, we count the frequency of each 3-gram chain in the seed program as a dimension in the feature vector.
\end{itemize}
As shown in Table~\ref{tab:techniques}, the methods for initial seed selection using the above three features are referred to as \tfidfset{}, \astset{}, and \cfgset{}.

\textbf{Code semantic extraction.} Recent studies have demonstrated the efficacy of leveraging pre-trained code representation models in various software engineering tasks~\cite{downtask1,downtask2,downtask3,downtask4}. 
Therefore, employing pre-trained models to obtain program feature vectors is intuitive. 
Particularly, in the study of JVM testing~\cite{vect}, the code representation models (CodeBERT~\cite{codebert}, InferCode~\cite{infercode}, codeT5~\cite{codet5}, and PLBART~\cite{plbart}) play a pivotal role in semantic vectorization.
Encouraged by this, we also adopt these four pre-trained models, aiming to explore their effectiveness in code representation for initial seed selection tasks. 
To do so, we use open-source pre-trained models directly for zero-shot inference of the code representation vectors from each seed program. 
If the seed program is too long, it is then divided into slices, and the code representation vectors for each slice are averaged. 
As shown in Table~\ref{tab:techniques}, the methods for initial seed selection using features extracted from these four pre-trained models are referred to as \codebertset{}, \infercodeset{}, \codetset{}, and \plbartset{}.

\textbf{Corpus subset selection.}
After feature extraction, \feabase{} sorts all seed programs and selects the top-k percent to form the corpus subset. 
The greedy algorithm is applicable to sortable data, e.g., coverage information used by \covbase{} and the number of detected bugs used by \fuzzbase{}. 
However, \feabase{} represents program features as vectors for selection, which are unsortable, so the greedy strategy is not applicable.
Some techniques built on the concept of adaptive random testing (ART)~\cite{art1,art2,art3} argue that uniformly distributed test cases are more likely to detect bugs with fewer test cases than ordinary random testing.
Furthest Point Sampling~\cite{fps} (FPS) is widely used in existing work to select uniformly distributed test cases~\cite{fpstask2,fpstask1}.
Following this, we employ FPS to sort the seed programs in the corpus by measuring distance between vectors, as seed programs with similar program features may reveal similar bugs.
The central concept of FPS is to prioritize selecting points that are farthest away from the centroid of the already selected points. 
This strategy ensures that the selected points exhibit distinct features, enabling the prioritization of more unique points in the process.

% \jj{how to measure the distance between two programs? We need to mention we obtain feature vectors (how?) from feature extraction, and then use some specific formulae to measure the distance?}

Give a set of initial seeds denoted as $\mathcal S=\{s_{1},s_{2},.. ,s_{n}\}$ (n refers to the number of seeds) and the budget of initial seed selection denoted as $k$, the target of \feabase{} is to select $k$ percent of the seed programs from $\mathcal S$ as a corpus subset. Specifically, \feabase{} first needs to extract the feature vectors for each seed program, denoted as $\mathcal V=\{v_{1},v_{2},... ,v_{n}\}$ and $v_{i}=f(s_{i})$, where $f$ refers to the feature extraction method and $f(s_{i})$ refers to the feature vector of $s_{i}$. Then, \feabase{} sorts the seed programs using FPS.  To ensure stability, we first compute the feature centroid of all seed programs as shown in Formula~\ref{equ:centroid}. 
\begin{equation}
\label{equ:centroid}
     C=(c_1,c_2,...,c_m)=\frac{1}{n} \sum_{i=1}^{n} f(s_i)
\end{equation}
Where $c_{i}$ refers to one dimension of a feature vector, $m$ refers to the number of dimensions. Then, \feabase{} selects the point farthest from the centroid as the first seed program shown as in Formula~\ref{equ:firstPoint}.
\begin{equation}
\label{equ:firstPoint}
    t_1 = \arg\max_{s_i \in \mathcal S} dist(f(s_i),C)
\end{equation}
Where $t_1$ refers to the first selected point and $dist(v_i,C)$ calculates the Euclidean distance between feature vectors. We use $\mathcal T$ to represent the set of seeds that have been selected. For each seed $s_i\in \mathcal S-\mathcal T$, we need to compute its minimum distance from the selected seeds in the set $\mathcal T$ and select the seed with the maximum distance shown as Formula~\ref{equ:maxmindistance}. 
\begin{equation}
\label{equ:maxmindistance}
    t_{next} = \arg\max_{s_i\in \mathcal S-\mathcal T}\min_{t_j \in \mathcal T} dist(f(s_i),f(t_j))
\end{equation}
This iterative process continues until all seed programs are included in $\mathcal T$, indicating the completion of sorting. 
Lastly, we exclusively select the top-k percent of seed programs from $\mathcal T$ to compose the corpus subset.
% Finally, we select the seed with maximum distance as the next shown as Formula~\ref{equ:maxdistance}
% Specifically, \feabase{} calculates the euclidean distance between the feature vectors 
% first extracts the features of each seed program, sorts the seed programs using FPS, and finally selects the top $k$ percent of the seed programs as the corpus subset. Specifically, 
% Algorithm~\ref{algo:corpusreduction} formally illustrates the initial seed selection process of \feabase{}. The first step is to extract features for each seed program (Lines 2-5), then use FPS to sort the initial seed in corpus (Line 6), and finally select the top k percent of the seed program as the corpus subset (Lines 7,8). To ensure the stability of \feabase{}, we first compute the feature centroid of all seed programs and select the point farthest from the centroid as the first seed program (Lines 10,11), incorporating it into the {\tt selectedList} (Line 12). This method of initial point selection guarantees higher specificity. Subsequently, we compute the feature centroid of all seed programs in {\tt selectedList}, choose the seed program farthest from the centroid, and add it to {\tt selectedList} (Lines 16-18). This iterative process continues until all seed programs are included in {\tt selectedList} (Line 15), indicating the completion of sorting. 
% Lastly, we exclusively select the top k percent of seed programs from {\tt sortedList} to compose the corpus subset (Lines 7,8).

\section{Evaluation Design}
\label{sec:evaluationdesign}

% \subsection{Research Questions}
% We conduct a large-scale evaluation to empirically analyze the effect of corpus reduction techniques in JVM testing. Three research questions are formulated to guide the study:

% \begin{itemize}
%     \item \textbf{RQ1}: How do corpus reduction techniques perform in JVM fuzzing? 
%     \item \textbf{RQ2}: How effective is the open-source corpus in JVM fuzzing?
%     \item \textbf{RQ3}: Can open-source corpus help detect previously unknown bugs?
% \end{itemize}

\subsection{Studied JVMs}
In line with existing research focusing on JVM testing~\cite{vect}, we studied three popular JVMs: Hotspot, OpenJ9, and Bisheng JDK. 
Table~\ref{tab:jvms} shows the summary of studied JVMs with their versions. 
Unlike existing work~\cite{javatailor,vect}, we did not use OpenJDK12-14 as they are no longer maintained.
Alternatively, we added OpenJDK17 as a subject. 
Note that, for OpenJDK-8, each JVM includes both an older build and the latest build. 
This is because relatively older JVM builds often contain more bugs, which allows for a better evaluation of the fuzzer's effectiveness.
By comparing the outputs of older JVM builds to those of the latest builds, we can calculate the evaluation metric of unique inconsistencies.
Similarly, by examining the outputs of the latest builds, we can calculate the evaluation metric of unknown bugs.
Further details regarding these evaluation metrics are provided in Section~\ref{subsec:measurements}.

\begin{table}[t]
\caption{Studied JVM versions}
\label{tab:jvms}
\centering
\small
\setlength{\tabcolsep}{3.5mm}
\renewcommand{\arraystretch}{1.5}
\begin{spacing}{1.2}
\resizebox{.99\linewidth}{!}{
\begin{tabular}{ccr}
\toprule
\textbf{\begin{tabular}[c]{@{}c@{}}OpenJDK\\ Version\end{tabular}} &
  \textbf{\begin{tabular}[c]{@{}c@{}}JVM\\ Implementation\end{tabular}} &
  \multicolumn{1}{c}{\textbf{ Version}} \\ \midrule
\multirow{6}{*}{OpenJDK8}  & \multirow{2}{*}{HotSpot}     &  \cellcolor[HTML]{E8E8E8}build 25.0-b70    \\ \cline{3-3} 
                           &                              & build 25.402-b06                             \\ \cmidrule(l){2-3} 
                           & \multirow{2}{*}{OpenJ9}      & \cellcolor[HTML]{E8E8E8}build openj9-0.8.0     \\ \cline{3-3} 
                           &                              & build openj9-0.43.0                             \\ \cmidrule(l){2-3} 
                           & \multirow{2}{*}{Bisheng JDK} & \cellcolor[HTML]{E8E8E8}build 25.302-b13     \\ \cline{3-3} 
                           &                              & build 25.392-b12                             \\ 
                           \hline
\multirow{3}{*}{OpenJDK11} & HotSpot                  & build 11.0.22+7    \\ \cline{2-3} 
                           & OpenJ9                   & build openj9-0.43.0    \\ \cline{2-3} 
                           & Bisheng JDK              & build 11.0.21+12    \\ 
                           \hline
\multirow{3}{*}{OpenJDK17} & HotSpot                  & build 17.0.10+7    \\ \cline{2-3} 
                           & OpenJ9                   & build openj9-0.43.0    \\ \cline{2-3} 
                           & Bisheng JDK              & build 17.0.9+12    \\
                           \bottomrule
\multicolumn{3}{l}{Shadow represents the used old build of the corresponding JVM.}
\end{tabular}
}
\vspace{-.3cm}
\end{spacing}
\end{table}

\subsection{Studied Corpus}
Two types of corpus are studied: benchmark corpus and open-source corpus.
Table~\ref{tab:benchmarks} provides basic information about these corpus, where \textit{\#Size} represents the number of seed programs and \textit{\#Inst} represents the number of Jimple instructions. 

\textbf{Benchmark Corpus.}
To ensure fair evaluation, we used the same corpus as the prior works focusing on JVM testing~\cite{javatailor,vect}. 
Specifically, we filtered out those corpus that contained only one seed program, as they did not require seed selection.
Finally, we obtained two corpus that contained multiple seed programs (i.e., P1 and P2) for this study.
% \gtc{Specifically, six of the eight corpus used in the existing work contain only one seed program, making initial seed selection unfeasible. Therefore, in our study, we selected only two of them, namely Hotspot-tests and OpenJ9-tests (referred to as P1 and P2).}. 
% The selected two corpus contain numerous test programs (i.e., 469 and 586, respectively), which are historical bug-revealing test programs collected from the repositories of HotSpot and OpenJ9~\cite{javatailor}, respectively. 
The studied two corpus consist of historical bug-revealing test programs collected from the Hotspot and OpenJ9 repositories~\cite{javatailor}, containing 469 and 586 test programs, respectively.

\begin{figure}[t]
    \centering
    \includegraphics[width=.9\linewidth]
    {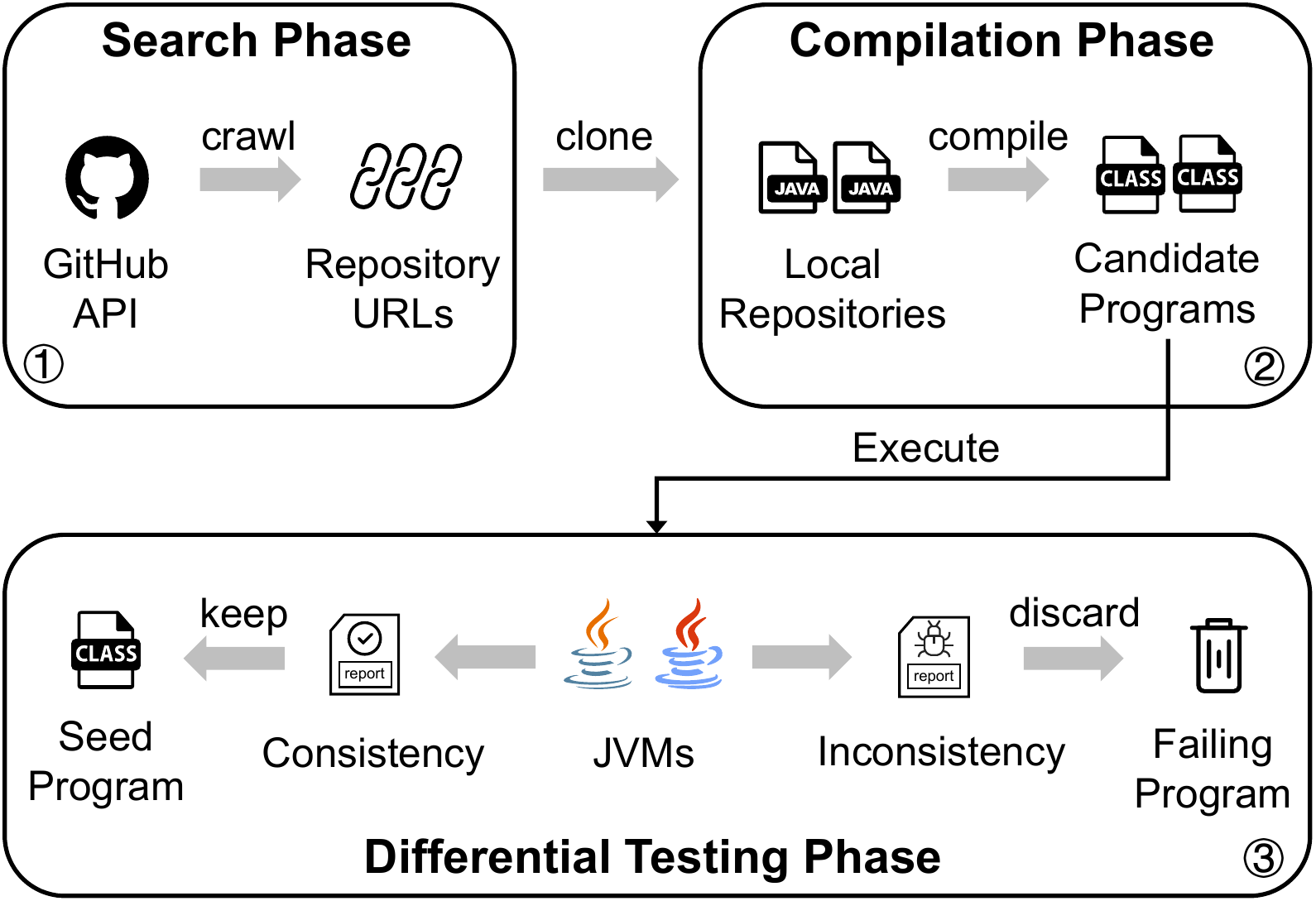}
    \caption{Automated collection of open-source corpus}
    \label{fig:corpus_collection}
    \vspace{-.5cm}
\end{figure}

\textbf{Open-Source Corpus.}
To enhance the corpus diversity for JVM testing, we propose a novel corpus sourced from the open-source community on GitHub, containing a larger set of test programs.
% The last corpus was sourced from the open-source community on GitHub, containing a larger set of test programs, but its effectiveness needs to be evaluated experimentally. 
% For the corpus collection, we chose the Github open-source community as the source of the seed program and implemented an automated collection script.
Figure~\ref{fig:corpus_collection} provides an overview of the corpus collection process. 
The process consists of three phases: search, compilation, and differential testing. 
% In this subsection, we will introduce the design and implementation of corpus collection.  

\begin{table}[t]
\caption{Studied corpus}
\label{tab:benchmarks}
\small
\centering
\begin{spacing}{1.2}
\resizebox{.8\linewidth}{!}{
\begin{tabular}{clrrc}
\toprule
\textbf{ID} & \multicolumn{1}{l}{\textbf{ProjectName}} 
            & \multicolumn{1}{c}{\textbf{\#Size}} 
            & \textbf{\#Inst} 
            & \textbf{Source}
             \\ \midrule
P1 & HotSpot-tests   & 469  & 104,797  & Hotspot    \\
P2 & OpenJ9-tests    & 586  & 105,212  & OpenJ9     \\  \midrule
P3 & CollectProject  & 942  & 127,798  & GitHub      \\ 
\bottomrule
\end{tabular}
}
\vspace{-.5cm}
\end{spacing}
\end{table}

\circled{1}~During the search phase, we collected URLs of projects from GitHub using the following criteria: the projects are written in Java, they have more than 100 stars, and the project structure is Maven. 
Specifically, we utilized the GitHub API~\cite{githubapi} as our search tool and set the aforementioned criteria to retrieve repository URLs. 
Because the GitHub API limits the return to fewer than 1,000 repositories, we segmented the star counts to ensure that the number of repositories returned each time stays within this limit.
After the search phase, we obtained a total of 6,507 repository URLs.

\circled{2}~During the compilation phase, we first filtered out those repositories that either do not contain a \textit{pom.xml} file or lack a Main entry.
The presence of a \textit{pom.xml} file facilitates the automation of the compilation process, while the Main entry point is essential for obtaining seed programs. 
We successfully cloned 460 of the surviving repositories locally for compilation.
Specifically, we used the \textit{mvn package} command to compile, however, if repositories lacked dependencies and could not execute the command, we resorted to using \textit{javac} to compile the Main entry in the repository. 
This approach ensures that we fully utilize each repository, maximizing the potential for successful compilation. 
After the compilation phase, we filtered out 256 repositories that either could not be compiled or did not have a successfully complied Main entry. 
Finally, 204 repositories and 1,173 Main entries remained.
% \wang{how many projects were filtered out and how many remained during this phase? how many class. files?}

\circled{3}~During the differential testing phase, we employed older JVM builds for testing.
The specific versions of them can be found in our replication package.
One issue with older JVMs is that certain seed programs may directly trigger inconsistencies, which could potentially interfere with the effectiveness of evaluations.
To mitigate this issue, we performed an interference on the Main entry in each repository to verify whether any of the seed programs were executing properly. 
This extra step aids in ensuring that the seed programs are functioning as expected, enabling us to identify and address any discrepancies before evaluations. 
After the differential testing phase, we filtered out 50 repositories that did not obtain any consistent Main entry, leaving us with 154 repositories. 
In particular, we filtered out 231 inconsistent Main entry classes. 
As shown in Table~\ref{tab:benchmarks}, our proposed corpus (referred to as P3) finally consists of 942 seed programs and 127,798 Jimple instructions.
% Table \ref{tab:dtjvms} shows the specific versions of JVMs used in differential testing.
% \wang{thanks for addressing this. The table I think we could place it in the replication package. Also, how do we get 942 and 127,798 as shown in Table III.}

\subsection{Experimented Fuzzers}

% \jj{please don't mention the classification of JVM fuzzing techniques, which will limit the deiversity of the studied techniques explicitly.}
In this study, two seed-driven JVM fuzzers are selected: \textit{JavaTailor} and \textit{VECT}. 
These two fuzzers construct a test program by synthesizing various code snippets, i.e., putting various ingredients extracted from historical bug-revealing test programs into a new context provided by a given seed program. They are state-of-the-art and have outperformed other widely-studied JVM testing techniques as demonstrated by prior work~\cite{javatailor,vect}.
The detailed descriptions of these two fuzzers are presented in Section~\ref{sec:background}.

Note that we opted not to use the enhanced test oracle featured in VECT. 
This choice was made because we did not manually filter or modify the open-source seed programs.
Certain variables, like randomness and timestamps, can introduce non-determinism. Therefore, the Correcting Commit~\cite{correctingcommit} used for inconsistency deduplication may potentially introduce false positives, which could impact our results.
Therefore, this study employed the same test oracle as JavaTailor in the evaluation.
Considering the effectiveness of VECT, we adopted PLBART, as recommended in VECT, for semantic vectorization.

% Based on recommendations from related work, we used benchmark corpus P1 as the historical bug-revealing test project for extracting ingredients in JavaTailor and VECT \jj{this can be omitted}. 
% Both fuzzers' implementations initially had only one seed pool, and synthesized programs would be put to the seed pool during testing, awaiting higher-order synthesis. However, this implementation could potentially change the distribution of the corpus. To address this, we modified the implementation to ensure that each seed program has its own separate seed pool, thereby maintaining the distribution of the corpus throughout the fuzzing process \jj{is this adaptation critical? this may cause confusion.}. 
% \jj{this content is related to experimental process, which doesn't belong to this subsection of studied fuzzers.}.

\subsection{Measurement}
\label{subsec:measurements}

% We compare the performance of each initial seed selection method across three measurements:

\textit{\textbf{Time overhead:}}
To compare the time overheads of each initial seed selection method, we recorded the time required for three key phases: data collection, data processing, and corpus subset selection. 
Specifically, during the data collection phase, \covbase{} needs to collect coverage data from the JVM and \fuzzbase{} gathers the fuzzing results of each seed program.
On the other hand, \feabase{} directly utilizes program features, thus eliminating the need for data collection. 
During the data processing phase, \covbase{} constructs bitmaps to facilitate subsequent operations, \fuzzbase{} performs deduplication and counts the number of inconsistencies detected, and \feabase{} extracts program features and represents them as vectors for further analysis. 
During the corpus subset selection phase, either greedy strategies or Farthest Point Sampling (FPS) are employed for sorting, and the top-k percent of seed programs are selected.

\textit{\textbf{Number of unique inconsistencies:}}
In the differential testing experiments conducted on relatively older JVM builds, each studied method is likely to detect some inconsistencies (i.e., potential real bugs, false positives, and duplicates) during the same testing time. 
Therefore, we executed another round of differential testing employing the latest JVM builds to verify whether the inconsistencies remain.
If the inconsistency disappears, it suggests that a known bug has been fixed; otherwise, we performed manual analysis to determine whether the inconsistency is due to a potential bug. 
Additionally, some inconsistencies may be duplicated due to the same bug, hence we further de-duplicated them based on the crash messages, which is the most commonly used automatic method in existing work~\cite{vect}. 
We used the number of inconsistencies after de-duplication as the number of unique inconsistencies. 

% \gtc{rewrite: Following existing work, we conducted two differential testing experiments on relatively older and the latest JVM versions. If discrepancies are only detected in the older JVM version, it implies that a known bug has been fixed; otherwise, we conduct manual analysis to determine if the discrepancy is due to a potential bug. After that, relying on the crash messages is the most commonly used automatic de-duplication method in existing work, hence we further de-duplicated them based on the crash messages. Although VECT proposes an enhanced test oracle, the open-source corpora collected lack analysis and tuning, making the use of the Checksum monitoring variable infeasible. Consequently, our study only records inconsistencies caused by crash and exception, without incorporating the enhanced test oracle proposed by VECT.}
% \wang{This justification is similar to the ones introduced in VECT. Do we need it or simplify it?}

\textit{\textbf{Number of previously unknown bugs:}}
In the differential-testing experiments conducted on newer JVM builds, each studied technique is likely to detect some inconsistencies.
In RQ3, we chose the more efficient VECT to conduct differential testing experiments on the latest versions of JVM.
Existing research demonstrates that while VECT and JavaTailor share the same exploration space, VECT exhibits comparatively greater exploration efficiency~\cite{vect}.  
To verify the authenticity of unknown bugs, we manually analyzed each inconsistency to determine whether a discrepancy is a real bug.
Then, we created bug reports for test programs and submitted them to the corresponding repositories. 
The number of unknown bugs is measured based on feedback from developers.

\subsection{Implementation and Environment}
\label{subsubsec:environment}

We implemented all initial seed selection methods in Java.
The extraction and analysis of AST and CFG relied on APIs provided by JDT and Soot. 
We utilized the pre-trained code representation models from existing work\cite{vect}, using their default parameters without any fine-tuning.
Extraction of code representation and corpus collection is implemented in Python.
To mitigate the effect of randomness, all experimental results related to inconsistencies were averaged over five repetitions of experiments, with each experiment set to run for 24 hours.
Experiments were conducted on a server with two dodeca-core CPUs Intel(R) Xe on(R) Silver 4214 CPU @ 2.20GHz and 251GB RAM, running Ubuntu 18.04.4 LTS (64-bit). 

\section{Result Analysis}
\begin{table}[t]
\caption{Comparison results of different methods in terms of the number of unique inconsistencies}
\small
\label{tab:unique_inconsistency}
\begin{spacing}{1.2}
\centering
\resizebox{0.85\linewidth}{!}{
\begin{tabular}{crcccccc}
\toprule

\multicolumn{1}{c}{} &
\multicolumn{1}{c}{} &
    \multicolumn{3}{|c}{JavaTailor} &
    \multicolumn{3}{|c}{VECT} \\ 
    \midrule

\multirow{13}{*}{\textbf{P1}} &
\multicolumn{1}{|l}{\fullset{}}  & 
    \multicolumn{3}{|c}{5.4} & 
    \multicolumn{3}{|c}{9.0} \\
    \cmidrule(l){2-8} &

\multicolumn{1}{|c}{} &
    \multicolumn{1}{|c}{20\%} &
    \multicolumn{1}{c}{35\%} &
    \multicolumn{1}{c}{50\%} &
    \multicolumn{1}{|c}{20\%} &
    \multicolumn{1}{c}{35\%} &
    \multicolumn{1}{c}{50\%} \\
    \cmidrule(l){2-8} &

\multicolumn{1}{|l}{\randomset{}} & 
    \multicolumn{1}{|c}{5.0} & 
    \multicolumn{1}{c}{4.6} & 
    \multicolumn{1}{c}{5.4} & 
    
    \multicolumn{1}{|c}{5.6} & 
    \multicolumn{1}{c}{6.2} & 
    \multicolumn{1}{c}{7.4} \\  
    \cmidrule(l){2-8} &
    
\multicolumn{1}{|l}{\minset{}} & 
    \multicolumn{1}{|c}{\textbf{7.4}} & 
    \multicolumn{1}{c}{\textbf{6.0}} & 
    \multicolumn{1}{c}{\textbf{7.8}} & 
    
    \multicolumn{1}{|c}{7.2} & 
    \multicolumn{1}{c}{\cellcolor[HTML]{CCCCCC}\textbf{9.4}} & 
    \multicolumn{1}{c}{\cellcolor[HTML]{DDDDDD}8.4} \\ &
    
\multicolumn{1}{|l}{\peachset{}} & 
    \multicolumn{1}{|c}{5.0} & 
    \multicolumn{1}{c}{5.2} & 
    \multicolumn{1}{c}{\textbf{6.4}} & 
    
    \multicolumn{1}{|c}{5.4} & 
    \multicolumn{1}{c}{6.6} & 
    \multicolumn{1}{c}{7.4} \\ 
    \cmidrule(l){2-8} &
    
\multicolumn{1}{|l}{\hotset{}} & 
    \multicolumn{1}{|c}{\textbf{6.4}} & 
    \multicolumn{1}{c}{\textbf{7.2}} & 
    \multicolumn{1}{c}{\textbf{8.0}} & 
    
    \multicolumn{1}{|c}{\cellcolor[HTML]{CCCCCC}\textbf{9.4}} & 
    \multicolumn{1}{c}{\cellcolor[HTML]{DDDDDD}8.6} & 
    \multicolumn{1}{c}{7.8} \\ 
    \cmidrule(l){2-8} &
    
\multicolumn{1}{|l}{\tfidfset{}} & 
    \multicolumn{1}{|c}{\textbf{7.2}} & 
    \multicolumn{1}{c}{\textbf{8.4}} & 
    \multicolumn{1}{c}{\cellcolor[HTML]{CCCCCC}\textbf{9.6}} & 
    
    \multicolumn{1}{|c}{6.6} & 
    \multicolumn{1}{c}{7.6} & 
    \multicolumn{1}{c}{7.6} \\ &
    
\multicolumn{1}{|l}{\astset{}} & 
    \multicolumn{1}{|c}{\cellcolor[HTML]{CCCCCC}\textbf{8.6}} & 
    \multicolumn{1}{c}{\cellcolor[HTML]{CCCCCC}\textbf{9.0}} & 
    \multicolumn{1}{c}{\cellcolor[HTML]{DDDDDD}\textbf{9.4}} & 
    
    \multicolumn{1}{|c}{\cellcolor[HTML]{DDDDDD}8.6} & 
    \multicolumn{1}{c}{\cellcolor[HTML]{DDDDDD}8.6} & 
    \multicolumn{1}{c}{\cellcolor[HTML]{CCCCCC}\textbf{10.0}} \\ &
    
\multicolumn{1}{|l}{\cfgset{}} & 
    \multicolumn{1}{|c}{\cellcolor[HTML]{BBBBBB}\textbf{12.6}} & 
    \multicolumn{1}{c}{\cellcolor[HTML]{BBBBBB}\textbf{13.0}} & 
    \multicolumn{1}{c}{\cellcolor[HTML]{BBBBBB}\textbf{12.0}} &
    
    \multicolumn{1}{|c}{\cellcolor[HTML]{BBBBBB}\textbf{12.8}} & 
    \multicolumn{1}{c}{\cellcolor[HTML]{BBBBBB}\textbf{14.8}} & 
    \multicolumn{1}{c}{\cellcolor[HTML]{BBBBBB}\textbf{16.2}} \\ 
    \cmidrule(l){2-8} &
    
\multicolumn{1}{|l}{\codebertset{}} & 
    \multicolumn{1}{|c}{\textbf{7.6}} & 
    \multicolumn{1}{c}{\textbf{7.0}} & 
    \multicolumn{1}{c}{\textbf{8.6}} &  
    
    \multicolumn{1}{|c}{6.2} & 
    \multicolumn{1}{c}{7.4} & 
    \multicolumn{1}{c}{8.2} \\ &
    
\multicolumn{1}{|l}{\infercodeset{}} & 
    \multicolumn{1}{|c}{\textbf{7.2}} & 
    \multicolumn{1}{c}{\cellcolor[HTML]{DDDDDD}\textbf{8.8}} & 
    \multicolumn{1}{c}{\cellcolor[HTML]{DDDDDD}\textbf{9.4}} &
    
    \multicolumn{1}{|c}{7.8} & 
    \multicolumn{1}{c}{7.8} & 
    \multicolumn{1}{c}{\cellcolor[HTML]{DDDDDD}8.4} \\ &
    
\multicolumn{1}{|l}{\plbartset{}} & 
    \multicolumn{1}{|c}{\textbf{7.4}} & 
    \multicolumn{1}{c}{\textbf{8.2}} & 
    \multicolumn{1}{c}{\textbf{8.0}} & 
    
    \multicolumn{1}{|c}{7.6} & 
    \multicolumn{1}{c}{7.8} & 
    \multicolumn{1}{c}{7.2} \\ &
    
\multicolumn{1}{|l}{\codetset{}} & 
    \multicolumn{1}{|c}{\cellcolor[HTML]{DDDDDD}\textbf{8.0}} & 
    \multicolumn{1}{c}{\textbf{8.0}} & 
    \multicolumn{1}{c}{\textbf{9.0}} &  
    
    \multicolumn{1}{|c}{7.2} & 
    \multicolumn{1}{c}{\cellcolor[HTML]{DDDDDD}8.6} & 
    \multicolumn{1}{c}{8.2} \\ \midrule
    
\multirow{13}{*}{\textbf{P2}} &
    \multicolumn{1}{|l}{\fullset{}}  & 
    \multicolumn{3}{|c}{6.4} & 
    \multicolumn{3}{|c}{7.8} \\
    \cmidrule(l){2-8} &

\multicolumn{1}{|c}{} &
    \multicolumn{1}{|c}{20\%} &
    \multicolumn{1}{c}{35\%} &
    \multicolumn{1}{c}{50\%} &
    \multicolumn{1}{|c}{20\%} &
    \multicolumn{1}{c}{35\%} &
    \multicolumn{1}{c}{50\%} \\
    \cmidrule(l){2-8} &

\multicolumn{1}{|l}{\randomset{}} & 
    
    \multicolumn{1}{|c}{6.2} & 
    \multicolumn{1}{c}{\textbf{7.4}} & 
    \multicolumn{1}{c}{5.8} & 
    
    \multicolumn{1}{|c}{4.0} & 
    \multicolumn{1}{c}{5.0} & 
    \multicolumn{1}{c}{5.4} \\  
    \cmidrule(l){2-8} &
    
\multicolumn{1}{|l}{\minset{}} & 

    \multicolumn{1}{|c}{\textbf{8.6}} & 
    \multicolumn{1}{c}{\cellcolor[HTML]{DDDDDD}\textbf{8.2}} & 
    \multicolumn{1}{c}{\textbf{8.6}} & 
    
    \multicolumn{1}{|c}{\textbf{9.2}} & 
    \multicolumn{1}{c}{\cellcolor[HTML]{DDDDDD}\textbf{8.6}} & 
    \multicolumn{1}{c}{\textbf{9.2}} \\ &
    
\multicolumn{1}{|l}{\peachset{}} & 

    \multicolumn{1}{|c}{\cellcolor[HTML]{DDDDDD}\textbf{9.2}} & 
    \multicolumn{1}{c}{\textbf{7.4}} & 
    \multicolumn{1}{c}{\textbf{7.8}} & 
    
    \multicolumn{1}{|c}{\cellcolor[HTML]{DDDDDD}\textbf{9.4}} & 
    \multicolumn{1}{c}{\cellcolor[HTML]{CCCCCC}\textbf{9.4}} & 
    \multicolumn{1}{c}{\textbf{9.2}} \\ 
    \cmidrule(l){2-8} &
    
\multicolumn{1}{|l}{\hotset{}} & 

    \multicolumn{1}{|c}{\textbf{7.8}} & 
    \multicolumn{1}{c}{\textbf{7.2}} & 
    \multicolumn{1}{c}{\textbf{7.8}} & 
    
    \multicolumn{1}{|c}{\cellcolor[HTML]{CCCCCC}\textbf{9.6}} & 
    \multicolumn{1}{c}{\textbf{8.0}} & 
    \multicolumn{1}{c}{7.4} \\ 
    \cmidrule(l){2-8} &
    
\multicolumn{1}{|l}{\tfidfset{}} & 
    \multicolumn{1}{|c}{\textbf{7.0}} & 
    \multicolumn{1}{c}{\textbf{7.2}} & 
    \multicolumn{1}{c}{\textbf{6.8}} & 
    
    \multicolumn{1}{|c}{6.4} & 
    \multicolumn{1}{c}{6.0} & 
    \multicolumn{1}{c}{5.6} \\ &
    
\multicolumn{1}{|l}{\astset{}} & 

    \multicolumn{1}{|c}{\cellcolor[HTML]{CCCCCC}\textbf{11.8}} & 
    \multicolumn{1}{c}{\cellcolor[HTML]{CCCCCC}\textbf{9.2}} & 
    \multicolumn{1}{c}{\cellcolor[HTML]{CCCCCC}\textbf{11.6}} & 
    
    \multicolumn{1}{|c}{\textbf{9.0}} & 
    \multicolumn{1}{c}{\cellcolor[HTML]{CCCCCC}\textbf{9.4}} & 
    \multicolumn{1}{c}{\cellcolor[HTML]{CCCCCC}\textbf{10.2}} \\ &
    
\multicolumn{1}{|l}{\cfgset{}} & 

    \multicolumn{1}{|c}{\cellcolor[HTML]{BBBBBB}\textbf{15.8}} & 
    \multicolumn{1}{c}{\cellcolor[HTML]{BBBBBB}\textbf{15.0}} & 
    \multicolumn{1}{c}{\cellcolor[HTML]{BBBBBB}\textbf{17.2}} & 
    
    \multicolumn{1}{|c}{\cellcolor[HTML]{BBBBBB}\textbf{16.6}} & 
    \multicolumn{1}{c}{\cellcolor[HTML]{BBBBBB}\textbf{16.6}} & 
    \multicolumn{1}{c}{\cellcolor[HTML]{BBBBBB}\textbf{15.4}} \\ 
    \cmidrule(l){2-8} &
    
\multicolumn{1}{|l}{\codebertset{}} & 
    
    \multicolumn{1}{|c}{\textbf{6.6}} & 
    \multicolumn{1}{c}{\textbf{7.6}} & 
    \multicolumn{1}{c}{\textbf{8.2}} & 
    
    \multicolumn{1}{|c}{6.6} & 
    \multicolumn{1}{c}{6.8} & 
    \multicolumn{1}{c}{\textbf{8.8}} \\ &
    
\multicolumn{1}{|l}{\infercodeset{}} & 

    \multicolumn{1}{|c}{6.4} & 
    \multicolumn{1}{c}{\textbf{7.8}} & 
    \multicolumn{1}{c}{\textbf{6.6}} & 
    
    \multicolumn{1}{|c}{7.2} & 
    \multicolumn{1}{c}{5.4} & 
    \multicolumn{1}{c}{5.6} \\ &
    
\multicolumn{1}{|l}{\plbartset{}} & 

    \multicolumn{1}{|c}{6.2} & 
    \multicolumn{1}{c}{\textbf{7.0}} & 
    \multicolumn{1}{c}{\textbf{6.6}} & 
    
    \multicolumn{1}{|c}{6.2} & 
    \multicolumn{1}{c}{5.6} & 
    \multicolumn{1}{c}{5.4} \\ &
    
\multicolumn{1}{|l}{\codetset{}} & 

    \multicolumn{1}{|c}{6.4} & 
    \multicolumn{1}{c}{\textbf{7.4}} & 
    \multicolumn{1}{c}{\cellcolor[HTML]{DDDDDD}\textbf{8.8}} & 
    
    \multicolumn{1}{|c}{6.6} & 
    \multicolumn{1}{c}{\textbf{8.4}} & 
    \multicolumn{1}{c}{\cellcolor[HTML]{DDDDDD}\textbf{9.6}} \\ 
\bottomrule
\multicolumn{8}{l}{The top 3 performing methods are shaded. } \\
\multicolumn{8}{l}{Darker shades indicating better performance.}
\end{tabular}
}
\vspace{-.5cm}
\end{spacing}
\end{table}

\subsection{RQ1: Performance of Initial seed Selection}
% \jj{we should separate the presentation of experimental process and results.}
% We analyze the performance of initial seed selection methods in JVM fuzzing in terms of effectiveness and efficiency.

\textbf{Effectiveness of initial seed selection methods.} 
The budget of existing corpus minimization methods (i.e., PeachSet, MinSet, and OptiMin) applied to P1 and P2 is mainly between 20\% and 50\%.
Inspired by this, to reduce the fuzzing budgets, we initially selected subsets with budgets set at 20\%, 35\%, and 50\%. 
% 、jj{How to inspire?}
Then, we applied each studied initial seed selection method with three budgets on benchmark corpus P1 and P2.
% Specifically, the budget of the subset selected by existing work mainly between 20\% and 50\%, so we selected subsets with budgets set at 20\%, 35\%, and 50\%. 
JavaTailor and VECT conducted 24 hours of fuzzing on each subset, counting the number of detected unique inconsistencies. 
Table~\ref{tab:unique_inconsistency} shows the comparison results of each method in terms of unique inconsistencies in the differential testing. 
% To mitigate the influence of randomness, each experimental result represents the average of five repeated experiments. \Wang{this is particular for RQ1 or all the RQs? If the latter, please put it into Section IV.E}
The \fullset{} and \randomset{} in the second column represent the use of the entire set of initial seeds and the subset selected by random selection, serving as baselines for comparison. 
Results highlighted in bold indicate superior performance compared to the \fullset{}.
The top three performing methods are shaded, with darker shades indicating better performance.

% \jj{need better organization.}
% \jj{We should first mention that all selection methods outperform the full set within the given fuzzing time budget, confirming the necessity of initial seed selection. Then, we compare these methods with random selection, showing the importance of carefully designing selection strategies. Finally, we compared all of selection methods, showing the best one (CFG-based method).}
% \jj{Deep analysis is also important. We need to explain (1) why CFG is the best; (2) why the existing ones on traditional software perform worse; (3) why code representation performs worse than CFG and AST.}
\textit{Results.} From Table~\ref{tab:unique_inconsistency}, we can observe that 
(1) When compared to the full set (\fullset{}), all three types of initial seed selection methods can outperform the full set within the given fuzzing time.
This finding confirms the necessity of performing initial seed selection in JVM fuzzing.
% However, different budgets will affect the effectiveness of the corpus subset, indicating that evaluating the effectiveness of budget selection is significant.
(2) When compared to random selection (RandomSet), evaluation results suggest that randomly selecting a subset could have a negative impact, which underlines the importance of carefully designing selection strategies.
(3) When compared across the studied selection methods, \cfgset{} is the best-performing method.
It outperforms \fullset{} in all results, yielding between 1.42 and 2.69 times higher performance, and ranks in the top three in all column comparisons.
For example, in JavaTailor and P2, the number of unique inconsistencies for the three budgets is 15.8, 15.0, and 17.2 respectively, compared to 6.4 when using \fullset{}.
The possible reason is that semantic information effectively represents the feature of the program, thereby enhancing the effectiveness of sorting in FPS.
\astset{} follows behind \cfgset{}, as it extracts features using AST, which is likely to have a coarser granularity compared to CFG.
The pre-trained model methods, such as \codebertset{} and \infercodeset{}, tend to exhibit relatively unstable and inferior performance. 
This could potentially be attributed to these models not being fine-tuned specifically for initial seed selection tasks.
Meanwhile, we noted that while coverage-based methods excel in traditional program fuzzing, the extensive code space within the JVM poses a challenge. 
\peachset{} and \minset{} are likely unable to effectively leverage coverage for assessing the quality of seed programs, suggesting that a significant portion of the coverage may not be relevant for inconsistency detection.

\begin{tcolorbox}[colback=gray!5]
\textbf{Finding I}: The initial seed selection that utilizes control flow graph, \cfgset{}, outperforms all other studied methods, yielding between 1.42 and 2.69 times the results of \fullset{} within the same test time, with an average improvement of 101\%.
\end{tcolorbox}

\textbf{Effectiveness of budget selection.}
It is unknown whether the established budget sizes from the existing corpus minimization are the most effective for JVM fuzzing.
Particularly, we study a variety of initial seed selection methods and certain strategies may perform better given a specific budget.
% Existing work has not investigated the impact of initial seed selection method on JVM fuzzing under more wider budget range. 
Hence, we extend our investigation to include a wider budget range, spanning from 5\% to 95\%.
For our exploratory analysis, we specifically evaluated the effectiveness of these budgets on the P1 corpus and the VECT fuzzer.
Table~\ref{tab:budgets} shows the average results of five repeated experiments, with shadows representing the best performers in each row.

% \jj{We should highlight: (1) confirming the superiority of initial seed selection under various sizes compared to fullset and random selection;
% (2) demonstrating the best effectiveness comes from CFG-based method under various sizes (except 5\%);
% (3) demonstrating the overall best effectiveness across all sizes comes from CFG-based method, and the most proper size is around 50\%. Top-3 is 20\%-50\%, confirming the choice of our setting in the prior experiment.}
\textit{Results.} From Table~\ref{tab:budgets},  we can observe that 
(1) Regardless of the budget size, the studied initial seed selection methods relatively achieve superior performance when compared to the full set and the random selection.
(2) Most coverage-based, HotSet-based, and traditional feature-based methods tend to perform better with smaller budgets.
For instance, the number of inconsistencies for \minset{} and \cfgset{} ranges from 7.2 to 9.4 and 12.8 to 16.2 respectively when the budgets are between 20\% and 50\%, while the number of inconsistencies ranges from 6.3 to 8.0 and 8.6 to 11.2 respectively when the budgets are between 65\% and 95\%.
Conversely, the four pre-trained model methods are likely to achieve better performance with relatively large budgets.
For example, \infercodeset{}, \plbartset{}, and \codetset{} achieve the best performance when the budget is 65\%, with the number of inconsistencies being 9.4, 9,6, and 8,8, respectively.  
% Notably, when the budget is set to 50\% or 65\%, most techniques achieve their best performance.
The potential reason is that these models were not specifically optimized for tasks related to selecting initial seeds.
(3) \cfgset{} consistently delivers the best performance across most (except 5\%)  budget ranges, demonstrating its effectiveness in JVM fuzzing. 
When the budget for \cfgset{} is set between 20\% and 50\%, the top three results can be achieved, further confirming the initial budget setting choice in prior experiment.
 We suggest setting the \cfgset{} budget to 50\%, as it provides the best outcomes across various budgets.

\begin{tcolorbox}[colback=gray!5]
\textbf{Finding II}: 
% \jj{the observation should be updated according. we should highlight the effectiveness of CFG-based method across wider range of scales, and recommend the most proper range for practical use.}
% Most coverage-based, HotSet-based, and traditional feature-based methods tend to achieve better performance with relatively smaller budgets (i.e., 35\% and 50\%). Conversely, the methods based on pre-trained models are likely to perform better with larger budgets (i.e., 65\% and 80\%).
% Different kinds of initial seed selection methods tend to perform well under a certain budget range.
\cfgset{} continuously achieves the top performance across a wider range of budget settings.
Specifically, We recommend using \cfgset{} with a budget set of 50\% for initial seed selection, which has been found to achieve optimal performance.
\end{tcolorbox}

\begin{table}[t]
\centering
\caption{The number of inconsistencies with different budgets}
\label{tab:budgets}
\small
\begin{spacing}{1.2}
\resizebox{.95\linewidth}{!}{
\begin{tabular}{llllllll}
\toprule
  \textbf{Method} & 
  \textbf{5\%} &
  \textbf{20\%} &
  \textbf{35\%} &
  \textbf{50\%} &
  \textbf{65\%} &
  \textbf{80\%} &
  \textbf{95\%} \\ \midrule
\minset{} &
  7.7 &
  7.2 &
  \cellcolor[HTML]{CCCCCC}9.4 &
  8.4 &
  6.3 &
  8.0 &
  7.0 \\
\peachset{} &
  4.0 &
  5.4 &
  6.6 &
  7.4 &
  \cellcolor[HTML]{CCCCCC}8.7 &
  7.7 &
  8.0 \\ \midrule
\hotset{} &
  7.2 &
  \cellcolor[HTML]{CCCCCC}9.4 &
  8.6 &
  7.8 &
  9.2 &
  7.0 &
  5.8 \\ \midrule
\tfidfset{} &
  5.0 &
  6.6 &
  \cellcolor[HTML]{CCCCCC}7.6 &
  \cellcolor[HTML]{CCCCCC}7.6 &
  7.2 &
  7.4 &
  7.0 \\
\astset{} &
  3.2 &
  8.6 &
  8.6 &
  \cellcolor[HTML]{CCCCCC}10 &
  8.2 &
  6.4 &
  8.0 \\
\cfgset{} &
  6.0 &
  12.8 &
  14.8 &
  \cellcolor[HTML]{CCCCCC}16.2 &
  11.2 &
  9.4 &
  8.6 \\ \midrule
\codebertset{}  & 
  5.2 & 
  6.2 & 
  7.4 & 
  8.2 & 
  7.6 & 
  \cellcolor[HTML]{CCCCCC}9.4 & 
  7.8 \\
\infercodeset{} & 
  5.6 & 
  7.8 & 
  7.8 & 
  8.4 &
  \cellcolor[HTML]{CCCCCC}9.4 &
  8.6 &
  6.8                         \\
\plbartset{} &
  5.0 &
  7.6 &
  7.8 &
  7.2 &
  \cellcolor[HTML]{CCCCCC}9.6 &
  7.6 &
  8.2 \\
\codetset{} &
  7.4 &
  7.2 &
  8.6 &
  8.2 &
  \cellcolor[HTML]{CCCCCC}8.8 &
  8.6 &
  8.2 \\   \bottomrule
\end{tabular}
}
\end{spacing}
\vspace{-.2cm}
\end{table}

\textbf{Efficiency of initial seed selection methods.}
Table~\ref{tab:overhead} presents the time overhead of each initial seed selection method on corpus P1, including three phases: data collection, data processing, and initial seed selection. 
As similar trends are observed across different corpus, we only provided the results on P1.
Note that the budget setting and fuzzing techniques do not affect the time overhead, as we sort the seed program first.

\textit{Results.} As shown in the table, the total time overhead of all the traditional feature-based methods requires no more than 30 seconds to select corpus subsets (i.e., 26s, 4s, and 30s for \tfidfset{}, \astset{}, and \cfgset{}, respectively) which is significantly faster than the other methods.
For the methods based on pre-trained models, results show that most of the time is spent on data processing when compared to traditional ones.
% benefiting from the efficient processing of programs into traditional feature vectors. 
% Conversely, pre-trained model-based \feabase{} requires more time during the data processing phase for code representation, but still only requires no more than 15 minutes of time overhead. 
For the prefuzz-based method \fuzzbase{}, a considerable amount of time (i.e., 140,700s) is taken to collect data as it necessitates fuzzing each seed program for five minutes.
Coverage-based methods \covbase{} exhibit the highest time overhead across all three phases, taking over 60 hours in total as it collects the coverage of each seed program for more than 500 seconds.
% \jj{providing brief explanations.}
% This is primarily because \covbase{} needs to gather JVM coverage for each seed program during the data collection phase and process coverage files into bitmaps during the data processing phase. 
% Furthermore, the extensive coverage space also increases the time required for sorting coverage (increments).
% \wang{When we explain results, better to focus on statistics themselves, reducing the space for in-depth discussions (those discussions you could add them into Discussion/Implicaiton section. For example, you could refer to the statistics to explain the difference between different techniques during each phase. Then highlight the efficiency of \cfgset{}.}

\begin{tcolorbox}[colback=gray!5]
\textbf{Finding III}:
Traditional feature-based methods have a significantly smaller time overhead compared to other methods.
More specifically, \cfgset{} not only enhances the effectiveness of JVM fuzzing but also demonstrates its efficiency.
\end{tcolorbox}

\subsection{RQ2: Initial Seed Selection on Programs in the Wild}

\begin{table}[t]
  \centering
  \caption{Time overhead of each method in P1}
  \label{tab:overhead}
  \small
  \begin{spacing}{1.2}
  \resizebox{.95\linewidth}{!}{
  \begin{tabular}{lrrrr}
    \toprule
    \textbf{Method} 
    & \textbf{\begin{tabular}[c]{@{}c@{}}Data\\ Collection\end{tabular}} 
    & \textbf{\begin{tabular}[c]{@{}c@{}}Data\\ Processing\end{tabular}} 
    & \textbf{\begin{tabular}[c]{@{}c@{}}Seed\\ Selection\end{tabular}} 
    & \textbf{Sum}\\
    \midrule
    \minset{}       & 188,330 s & 36,779 s & 819 s &  225,928 s\\
    \peachset{}     & 188,330 s & 36,779 s & 41 s  &  225,150 s\\ \hline
    \hotset{}       & 140,700 s & 1 s      & 1 s   &  140,702 s\\ \hline
    \tfidfset{}     & 0 s       & 20 s     & 6 s   &  26 s\\ 
    \astset{}       & 0 s       & 3 s      & 1 s  &  4 s\\ 
    \cfgset{}       & 0 s       & 29 s     & 1 s   &  30 s\\ \hline
    \codebertset{}  & 0 s       & 395 s    & 1 s   &  396 s\\
    \infercodeset{} & 0 s       & 272 s    & 1 s   &  273 s\\
    \plbartset{}    & 0 s       & 477 s    & 1 s   &  478 s\\
    \codetset{}     & 0 s       & 878 s    & 1 s   &  879 s\\
    \bottomrule
  \end{tabular}
  }
  \end{spacing}
\vspace{-.3cm}
\end{table}

% To evaluate the effectiveness of initial seed selection on programs in the wild (open-source corpus), we carried out the following two analyses: 
% \jj{why we perform the two analyses and what's the logic between them?}
% (i) the overlap of detected unique inconsistencies and (ii) the performance at various budgets.

\begin{figure}[t]
  \centering
  \begin{subfigure}[b]{0.5\linewidth}
    \centering
    \includegraphics{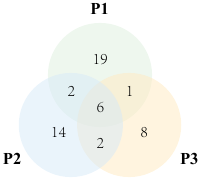}
    \caption{FullSet}
    \label{fig:fullsetoverlap}
  \end{subfigure}\hfill
  \begin{subfigure}[b]{0.5\linewidth}
    \centering
    \includegraphics{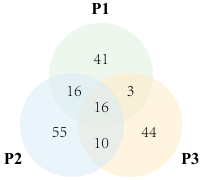}
    \caption{\cfgset{} with 50\% budget}
    \label{fig:cfgsetoverlap}
  \end{subfigure}
  \caption{Inconsistency overlap}
  \label{fig:overlap}
  \vspace{-.4cm}
\end{figure}

To evaluate the effectiveness of initial seed selection on programs in the wild, we first confirm the importance of the collected open-source corpus (P3) by analyzing the overlap of unique inconsistencies detected across the studied corpus (i.e., P1, P2, and P3). 
Then, we validate the generalizability of initial seed selection by repeating the experiments of RQ1 on the P3 corpus.
For the first analysis, we conducted fuzzing on \fullset{} by employing the same experimental settings from RQ1. 
For each corpus, we collated all inconsistencies detected by two fuzzers over five repeated runs, and then eliminated duplicates based on crash messages. 
For the second analysis, we applied each studied initial seed selection method on P3 with a wider budget range, i.e., spanning from 5\% to 95\%. 
We also performed an overlap analysis for the best-performing method.
The Venn diagrams in Figure~\ref{fig:overlap} show the results of the inconsistency overlap analysis, while Table~\ref{tab:p3_unique_inconsistency} records the results of the performance at various budgets. 

\textit{Results.} From Figure~\ref{fig:fullsetoverlap}, we found that each studied corpus can detect some unique inconsistencies with the full set, i.e., 19, 14, 8 for P1, P2, and P3, respectively. 
This finding suggests that the open-source corpus and the existing benchmark corpus complement each other.
In addition, the results show that P3 detected a total of 17 unique inconsistencies, while P1 and P2 were able to detect 28 and 24 inconsistencies separately.
This indicates that the open-source corpus is less effective in detecting unique inconsistencies and the quality of its seed program is relatively lower.

% i.e., 41, 55, 41 for P1, P2, and P3, respectively. Specifically, applying initial seed selection on P3 leads to a significant improvement in effectiveness, i.e., 73, 17 for \cfgset{} and FullSet, respectively. 
% This further demonstrates the effectiveness and generality of \cfgset{} in improving JVM fuzzing performance, especially when applied to open-source corpus where the improvement is more significant.

% To further analyse the effectiveness of corpus reduction applied in P3, we evaluated the performance of the selected corpus subsets from P3, with detailed results presented in Table~\ref{tab:unique_inconsistency}. 
\begin{table*}[htbp]
\caption{Comparison results on programs in the wild in terms of the number of unique inconsistencies}
\label{tab:p3_unique_inconsistency}
\small
\begin{spacing}{1.2}
\centering
\resizebox{0.8\linewidth}{!}{
\begin{tabular}{rrcccccccccccccc}
\toprule

\multicolumn{1}{c}{} &
\multicolumn{1}{c}{} &
    \multicolumn{7}{|c}{JavaTailor} &
    \multicolumn{7}{|c}{VECT} \\ \midrule

\multirow{13}{*}{\textbf{P3}} &
\multicolumn{1}{|l}{\fullset{}}  & 
    \multicolumn{7}{|c}{5.0} & 
    \multicolumn{7}{|c}{4.8} \\
    \cmidrule(l){2-16} &

\multicolumn{1}{|c}{} &
    \multicolumn{1}{|c}{5\%} &
    \multicolumn{1}{c}{20\%} &
    \multicolumn{1}{c}{35\%} &
    \multicolumn{1}{c}{50\%} &
    \multicolumn{1}{c}{65\%} &
    \multicolumn{1}{c}{80\%} &
    \multicolumn{1}{c}{95\%} &
    \multicolumn{1}{|c}{5\%} &
    \multicolumn{1}{c}{20\%} &
    \multicolumn{1}{c}{35\%} &
    \multicolumn{1}{c}{50\%} &
    \multicolumn{1}{c}{65\%} &
    \multicolumn{1}{c}{80\%} &
    \multicolumn{1}{c}{95\%} \\
\cmidrule(l){2-16} &

\multicolumn{1}{|l}{\randomset{}} & 
    \multicolumn{1}{|c}{4.3} & 
    \multicolumn{1}{c}{\textbf{5.2}} & 
    \multicolumn{1}{c}{4.4} & 
    \multicolumn{1}{c}{\textbf{5.6}} & 
    \multicolumn{1}{c}{5.0} & 
    \multicolumn{1}{c}{4.8} & 
    \multicolumn{1}{c}{\textbf{5.3}} & 

    \multicolumn{1}{|c}{4.2} & 
    \multicolumn{1}{c}{4.0} & 
    \multicolumn{1}{c}{3.8} & 
    \multicolumn{1}{c}{4.2}& 
    \multicolumn{1}{c}{\textbf{5.3}} & 
    \multicolumn{1}{c}{\textbf{5.0}} & 
    \multicolumn{1}{c}{\textbf{5.4}} \\  
    \cmidrule(l){2-16} &

\multicolumn{1}{|l}{\minset{}} & 
    \multicolumn{1}{|c}{5.0} & 
    \multicolumn{1}{c} {5.0} & 
    \multicolumn{1}{c} {\textbf{5.2}} & 
    \multicolumn{1}{c} {4.8} & 
    \multicolumn{1}{c} {\textbf{6.7}} & 
    \multicolumn{1}{c} {\textbf{5.3}} & 
    \multicolumn{1}{c} {5.0} & 

    \multicolumn{1}{|c}{\textbf{5.0}} & 
    \multicolumn{1}{c} {3.8} & 
    \multicolumn{1}{c} {4.8} & 
    \multicolumn{1}{c} {4.4} & 
    \multicolumn{1}{c} {\textbf{6.3}} & 
    \multicolumn{1}{c} {\textbf{5.8}} & 
    \multicolumn{1}{c} {\cellcolor[HTML]{DDDDDD}\textbf{5.3}} \\  &
    
\multicolumn{1}{|l}{\peachset{}} & 
    \multicolumn{1}{|c}{3.3} & 
    \multicolumn{1}{c} {4.2} & 
    \multicolumn{1}{c} {4.6} & 
    \multicolumn{1}{c} {4.4} & 
    \multicolumn{1}{c} {\textbf{7.3}} & 
    \multicolumn{1}{c} {4.7} & 
    \multicolumn{1}{c} {\textbf{5.3}} & 

    \multicolumn{1}{|c}{3.0} & 
    \multicolumn{1}{c} {3.6} & 
    \multicolumn{1}{c} {\textbf{5.2}} & 
    \multicolumn{1}{c} {3.8} & 
    \multicolumn{1}{c} {\textbf{5.8}} & 
    \multicolumn{1}{c} {\textbf{6.3}} & 
    \multicolumn{1}{c} {4.5} \\ 
    \cmidrule(l){2-16} &
    
\multicolumn{1}{|l}{\hotset{}} & 
    \multicolumn{1}{|c}{\textbf{7.0}} & 
    \multicolumn{1}{c} {\textbf{7.6}} & 
    \multicolumn{1}{c} {\textbf{5.8}} & 
    \multicolumn{1}{c} {\textbf{5.4}} & 
    \multicolumn{1}{c} {\textbf{7.0}} & 
    \multicolumn{1}{c} {\textbf{5.7}} & 
    \multicolumn{1}{c} {4.8} & 

    \multicolumn{1}{|c}{4.7} & 
    \multicolumn{1}{c} {\cellcolor[HTML]{DDDDDD}\textbf{6.8}} & 
    \multicolumn{1}{c} {\cellcolor[HTML]{CCCCCC}\textbf{7.0}} & 
    \multicolumn{1}{c} {\cellcolor[HTML]{CCCCCC}\textbf{7.6}} & 
    \multicolumn{1}{c} {\cellcolor[HTML]{CCCCCC}\textbf{9.0}} & 
    \multicolumn{1}{c} {\textbf{6.7}} & 
    \multicolumn{1}{c} {\textbf{5.0}} \\ 
    \cmidrule(l){2-16} &
    
\multicolumn{1}{|l}{\tfidfset{}} & 
    \multicolumn{1}{|c}{4.3} & 
    \multicolumn{1}{c} {\textbf{6.8}} & 
    \multicolumn{1}{c} {\textbf{5.6}} & 
    \multicolumn{1}{c} {\textbf{6.8}} & 
    \multicolumn{1}{c} {\cellcolor[HTML]{CCCCCC}\textbf{9.0}} & 
    \multicolumn{1}{c} {\textbf{7.0}} & 
    \multicolumn{1}{c} {\textbf{5.5}} & 

    \multicolumn{1}{|c}{3.3} & 
    \multicolumn{1}{c} {4.8} & 
    \multicolumn{1}{c} {\textbf{5.2}} & 
    \multicolumn{1}{c} {\textbf{6.8}} & 
    \multicolumn{1}{c} {\textbf{7.3}} & 
    \multicolumn{1}{c} {\textbf{6.5}} & 
    \multicolumn{1}{c} {\cellcolor[HTML]{CCCCCC}\textbf{5.5}} \\ &
    
\multicolumn{1}{|l}{\astset{}} & 
    \multicolumn{1}{|c}{3.3} & 
    \multicolumn{1}{c} {\textbf{6.0}} & 
    \multicolumn{1}{c} {\cellcolor[HTML]{CCCCCC}\textbf{7.8}} & 
    \multicolumn{1}{c} {\textbf{7.0}} & 
    \multicolumn{1}{c} {\textbf{8.3}} & 
    \multicolumn{1}{c} {\textbf{6.3}} & 
    \multicolumn{1}{c} {\cellcolor[HTML]{BBBBBB}\textbf{6.0}} & 

    \multicolumn{1}{|c}{3.0} & 
    \multicolumn{1}{c} {\textbf{6.0}} & 
    \multicolumn{1}{c} {\cellcolor[HTML]{DDDDDD}\textbf{6.6}} & 
    \multicolumn{1}{c} {\cellcolor[HTML]{DDDDDD}\textbf{7.0}} & 
    \multicolumn{1}{c} {\textbf{5.5}} & 
    \multicolumn{1}{c} {\cellcolor[HTML]{DDDDDD}\textbf{6.8}} & 
    \multicolumn{1}{c} {\cellcolor[HTML]{BBBBBB}\textbf{5.8}} \\ &
    
\multicolumn{1}{|l}{\cfgset{}} & 
    \multicolumn{1}{|c}{\cellcolor[HTML]{DDDDDD}\textbf{8.0}} & 
    \multicolumn{1}{c} {\cellcolor[HTML]{CCCCCC}\textbf{8.8}} & 
    \multicolumn{1}{c} {\cellcolor[HTML]{DDDDDD}\textbf{7.4}} & 
    \multicolumn{1}{c} {\cellcolor[HTML]{BBBBBB}\textbf{10.4}} & 
    \multicolumn{1}{c} {\cellcolor[HTML]{BBBBBB}\textbf{10.0}} & 
    \multicolumn{1}{c} {\cellcolor[HTML]{BBBBBB}\textbf{8.4}} & 
    \multicolumn{1}{c} {\cellcolor[HTML]{DDDDDD}\textbf{5.6}} & 

    \multicolumn{1}{|c}{\cellcolor[HTML]{DDDDDD}\textbf{5.2}} & 
    \multicolumn{1}{c} {\cellcolor[HTML]{CCCCCC}\textbf{7.4}} & 
    \multicolumn{1}{c} {\cellcolor[HTML]{DDDDDD}\textbf{6.6}} & 
    \multicolumn{1}{c} {\cellcolor[HTML]{BBBBBB}\textbf{9.6}} & 
    \multicolumn{1}{c} {\cellcolor[HTML]{BBBBBB}\textbf{9.6}} & 
    \multicolumn{1}{c} {\cellcolor[HTML]{CCCCCC}\textbf{7.4}} & 
    \multicolumn{1}{c} {\cellcolor[HTML]{CCCCCC}\textbf{5.5}} \\ 
    \cmidrule(l){2-16} &
    
\multicolumn{1}{|l}{\codebertset{}} & 
    \multicolumn{1}{|c}{\textbf{7.5}} & 
    \multicolumn{1}{c} {\cellcolor[HTML]{DDDDDD}\textbf{7.8}} & 
    \multicolumn{1}{c} {\textbf{7.0}} & 
    \multicolumn{1}{c} {\cellcolor[HTML]{DDDDDD}\textbf{7.6}} & 
    \multicolumn{1}{c} {\textbf{8.5}} & 
    \multicolumn{1}{c} {\textbf{7.0}} & 
    \multicolumn{1}{c} {5.0} & 

    \multicolumn{1}{|c}{4.7} & 
    \multicolumn{1}{c} {\textbf{6.2}} & 
    \multicolumn{1}{c} {\textbf{6.0}} & 
    \multicolumn{1}{c} {\textbf{6.2}} & 
    \multicolumn{1}{c} {\cellcolor[HTML]{DDDDDD}\textbf{7.8}} & 
    \multicolumn{1}{c} {\cellcolor[HTML]{CCCCCC}\textbf{7.3}} &  
    \multicolumn{1}{c} {\cellcolor[HTML]{DDDDDD}\textbf{5.3}} \\ &
    
\multicolumn{1}{|l}{\infercodeset{}} & 
    \multicolumn{1}{|c}{\cellcolor[HTML]{CCCCCC}\textbf{8.6}} & 
    \multicolumn{1}{c} {\cellcolor[HTML]{BBBBBB}\textbf{10.2}} & 
    \multicolumn{1}{c} {\cellcolor[HTML]{BBBBBB}\textbf{8.8}} & 
    \multicolumn{1}{c} {\textbf{7.4}} & 
    \multicolumn{1}{c} {\textbf{8.5}} & 
    \multicolumn{1}{c} {\cellcolor[HTML]{CCCCCC}\textbf{7.8}} & 
    \multicolumn{1}{c} {\cellcolor[HTML]{CCCCCC}\textbf{5.8}} & 

    \multicolumn{1}{|c}{\cellcolor[HTML]{BBBBBB}\textbf{6.8}} & 
    \multicolumn{1}{c} {\cellcolor[HTML]{BBBBBB}\textbf{9.2}} & 
    \multicolumn{1}{c} {\cellcolor[HTML]{BBBBBB}\textbf{8.0}} & 
    \multicolumn{1}{c} {4.2} & 
    \multicolumn{1}{c} {\textbf{7.5}} & 
    \multicolumn{1}{c} {\cellcolor[HTML]{CCCCCC}\textbf{7.3}} & 
    \multicolumn{1}{c} {\textbf{5.0}} \\ &
    
\multicolumn{1}{|l}{\plbartset{}} & 
    \multicolumn{1}{|c}{4.3} & 
    \multicolumn{1}{c} {\textbf{6.2}} & 
    \multicolumn{1}{c} {\cellcolor[HTML]{DDDDDD}\textbf{7.4}} & 
    \multicolumn{1}{c} {4.8} & 
    \multicolumn{1}{c} {\textbf{8.0}} & 
    \multicolumn{1}{c} {\textbf{6.7}} & 
    \multicolumn{1}{c} {\textbf{5.3}} & 

    \multicolumn{1}{|c}{2.5} & 
    \multicolumn{1}{c} {3.6} & 
    \multicolumn{1}{c} {3.8} & 
    \multicolumn{1}{c} {2.4} & 
    \multicolumn{1}{c} {\textbf{7.3}} & 
    \multicolumn{1}{c} {\textbf{6.5}} & 
    \multicolumn{1}{c} {\textbf{5.0}} \\ &
    
\multicolumn{1}{|l}{\codetset{}} & 
    \multicolumn{1}{|c}{\cellcolor[HTML]{BBBBBB}\textbf{8.8}} & 
    \multicolumn{1}{c} {\textbf{5.4}} & 
    \multicolumn{1}{c} {\textbf{6.4}} & 
    \multicolumn{1}{c} {\cellcolor[HTML]{CCCCCC}\textbf{7.8}} & 
    \multicolumn{1}{c} {\cellcolor[HTML]{DDDDDD}\textbf{8.8}} & 
    \multicolumn{1}{c} {\cellcolor[HTML]{DDDDDD}\textbf{7.3}} & 
    \multicolumn{1}{c} {4.8} & 

    \multicolumn{1}{|c}{\cellcolor[HTML]{CCCCCC}\textbf{5.8}} & 
    \multicolumn{1}{c} {\textbf{6.2}} & 
    \multicolumn{1}{c} {\textbf{5.6}} & 
    \multicolumn{1}{c} {\textbf{6.4}} & 
    \multicolumn{1}{c} {\textbf{6.5}} & 
    \multicolumn{1}{c} {\textbf{5.5}} & 
    \multicolumn{1}{c} {4.8} \\  
    \bottomrule
\end{tabular}
}
\vspace{-.5cm}
\end{spacing}
\end{table*}
From Table~\ref{tab:p3_unique_inconsistency}, we observed that on the open-source corpus, \cfgset{} still emerges as the best selection method, consistently ranking in the top three, with an improvement of 8.3\% to 108\% compared to the full set. 
For example, when the budget is set at 50\%, the number of detected unique inconsistencies with \cfgset{} reaches 10.4 for JavaTailor and 9.6 for VECT.
Besides, selection methods relying on pre-trained models show substantial performance enhancements. 
A possible reason is that, similar to P3, pre-trained models also use GitHub as a data source, which could potentially enhance the representation of code semantics.
This further illustrates that the lack of fine-tuning leads to the poor performance of pre-trained model-based methods on P1 and P2.
% \jj{Interesting point. We can explain more by comparing the three benchmark at this point.}

% \jj{we may also report the overlap analysis after initial seed selection. Whether the uniqueness of P3 can be improved?}

From Figure~\ref{fig:cfgsetoverlap}, we noticed that the most effective seed selection method (\cfgset{} with a 50\% budget) allows P3 to identify a comparable number of unique inconsistencies as P1 and P2.
For instance, the number of unique inconsistencies detected with P1, P2, and P3 is 41, 55, and 44, respectively.
Moreover, the number of unique inconsistencies detected by P3 improves significantly when compared to the full set, i.e., 73 and 17 for \cfgset{} and FullSet, respectively. 
The above results suggest that the initial seed selection can enhance the effectiveness of programs in the wild and further improve the fuzzing capability.

\begin{tcolorbox}[colback=gray!5]
\textbf{Finding IV}:
Programs in the wild complement widely-studied initial seeds by detecting new JVM behaviors. 
Selection methods relying on pre-trained models achieve substantial performance improvements in the open-source corpus.
Nonetheless, \cfgset{} remains the best-performing method, with an improvement of 8.3\% to 108\%, and enhances the effectiveness of the open-source corpus.
\end{tcolorbox}

% \begin{figure}
%     \centering
%     \includegraphics[width=0.5\linewidth]{figs/overlap_of_FullSet.pdf}
%     \caption{Inconsistency overlap in \fullset{}}
%     \label{fig:overlap}
% \end{figure}

\subsection{RQ3: Detection of Previously Unknown Bugs}
\begin{figure}
    \centering
    \includegraphics[width=0.9\linewidth]{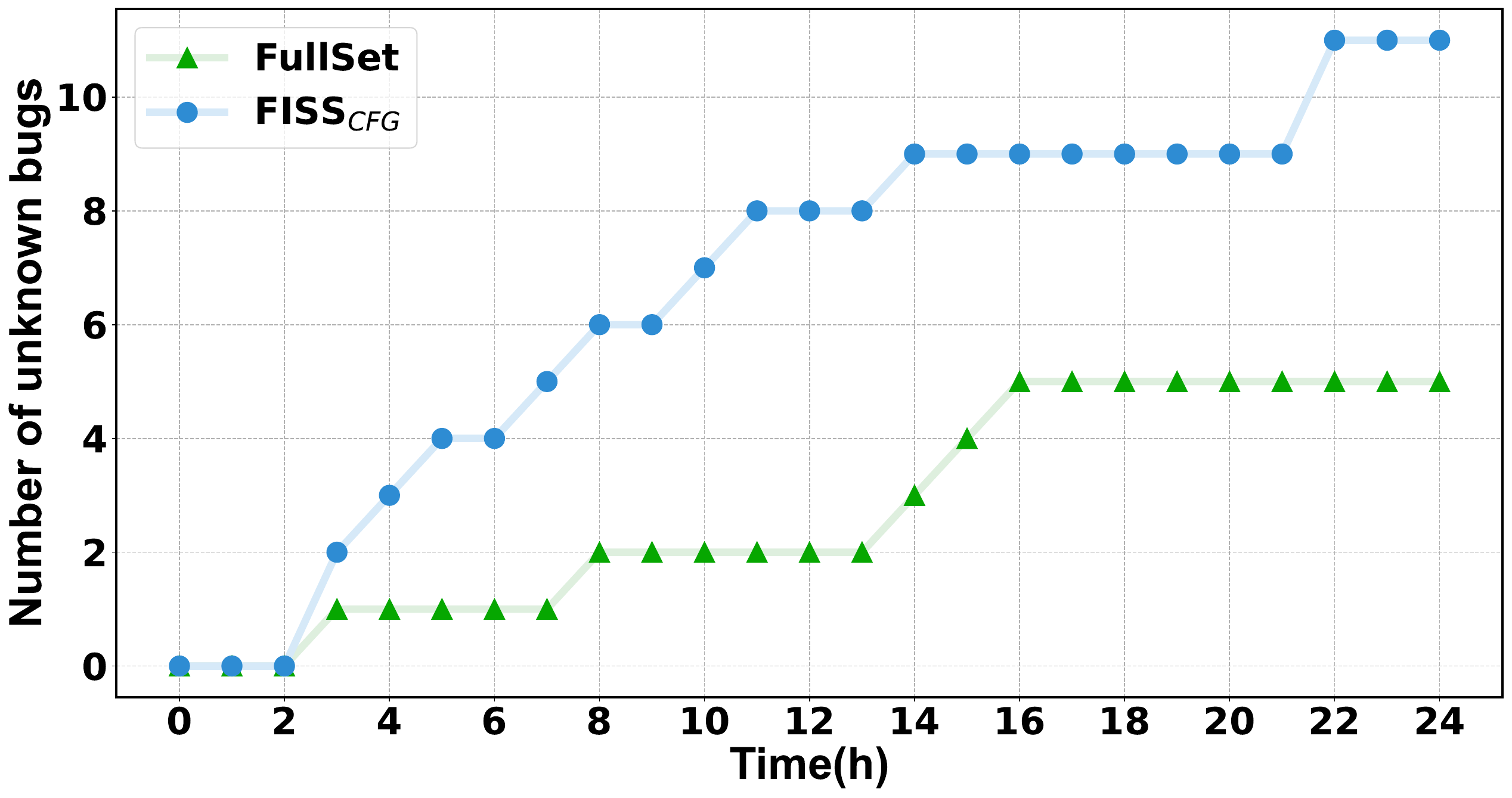}
    \caption{Trend of bugs detected by \cfgset{} and FullSet}
    \label{fig:bugtrend}
    \vspace{-.4cm}
\end{figure}

% \jj{I have told the new story to Tianchang. Please revise it.}
We further investigated whether the initial seed selection helps enhance JVM fuzzing techniques in detecting previously unknown bugs through the differential-testing experiments on the latest builds. 
Specifically, we applied VECT on all studied corpus, as it has been proven to be state-of-the-art~\cite{vect}.
For a fair comparison, we ran VECT 24 hours on both the FullSet and the subsets selected by each method with a 50\% budget.
We then submitted the detected inconsistencies on the latest builds to the corresponding developers to check whether it was a real bug.
% \gtc{Current research suggests that while JavaTailor can identify certain bugs, VECT not only detects these same bugs but also does so with greater exploration efficiency.}
% \gtc{Hence, we applied VECT to fuzzing each corpus on the latest builds and further manually investigated it to report the potential bug to the developers of JVMs.}
% \gtc{To fairly compare the ability of each method in detecting unknown bugs, VECT conducted 24 hours of fuzzing on both the FullSet and the subsets selected by each method with a 50\% budget.}
Table~\ref{table:bug_count} shows the total number of bugs discovered by each initial seed subset across the three studied corpus within the same testing period, and Figure~\ref{fig:bugtrend} shows the trend of bugs for FullSet and \cfgset{}.
% To compare the ability of \cfgset{} and FullSet to detect unknown bugs within the same time, we used all submitted bugs as ground truth. 
% VECT conducted 24 hours of fuzzing on both the FullSet and the subset selected by \cfgset{} with a 50\% budget. Figure~\ref{fig:bugtrend} shows the number of bugs discovered by the two sets of initial seeds within the same time.

\begin{table}[t]
\small
\centering
\caption{Confirmed or fixed unknown bugs}
\label{tab:unknown_bug}
\begin{spacing}{1.2}
\resizebox{.95\linewidth}{!}{
\begin{tabular}{lcccc}
\toprule
\textbf{Bug ID} & \textbf{JVM} & \textbf{\begin{tabular}[c]{@{}c@{}}Affected \\ OpenJDK\end{tabular}} & \textbf{Status} & \textbf{Corpus} \\
\midrule
\#15061     & OpenJ9      & 11      & fixed     & P1       \\
\#17247     & OpenJ9      & 17      & fixed     & P1       \\
\#17248     & OpenJ9      & 17      & fixed     & P1,P2,P3 \\
\#19014     & OpenJ9      & 11,17   & fixed     & P1       \\
\#19015     & OpenJ9      & 8,11,17 & fixed     & P2       \\
\#19016     & OpenJ9      & 8,11,17 & fixed     & P3       \\
\#19124     & OpenJ9      & 8,11,17 & confirmed & P1       \\
\#19125     & OpenJ9      & 8,11,17 & confirmed & P1,P2,P3 \\
\#19129     & OpenJ9      & 8       & fixed     & P1,P2    \\
\#19130     & OpenJ9      & 8,11,17 & confirmed & P1       \\
\#19132     & OpenJ9      & 17      & fixed     & P3       \\
\#19139     & OpenJ9      & 8       & confirmed & P1       \\
\#19140     & OpenJ9      & 8,11,17 & fixed     & P1       \\
\#19163     & OpenJ9      & 17      & confirmed & P1       \\
JDK-8326996 & Hotspot     & 11,17   & confirmed & P3       \\
JDK-8327011 & Hotspot     & 8       & confirmed & P3       \\
JDK-8327012 & Hotspot     & 17      & confirmed & P1,P2    \\ 
JDK-8328298 & Hotspot     & 11,17   & confirmed & P3    \\ 
\#I98GAP    & Bisheng JDK & 17      & confirmed & P1,P2    \\
\#I98GCO    & Bisheng JDK & 8       & confirmed & P3       \\
\#I98GD8    & Bisheng JDK & 11      & confirmed & P3       \\
\bottomrule
\end{tabular}
}
\end{spacing}
\vspace{-.3cm}
\end{table}

\textit{Results.} 
As shown in Talbe~\ref{table:bug_count}, \cfgset{} is the best-performing method across all the studied selection methods, yielding between 1.4 and 2.2 times higher performance.
Figure~\ref{fig:bugtrend} further shows the trend of the number of bugs detected by FullSet and \cfgset{} over time, where the x-axis represents the testing time, 
while the y-axis represents the number of unknown bugs detected by FullSet and \cfgset{} within the corresponding testing time.
As shown in Figure~\ref{fig:bugtrend}, \cfgset{} detected more unknown bugs than FullSet during the entire testing process, specifically 11 and 5 bugs respectively.
In the first 12 hours of fuzzing, \cfgset{} was able to detect unknown bugs more quickly. 
This is anticipated as \cfgset{} selects a subset with higher quality for the corpus, resulting in enhanced bug discovery capabilities.
Notably, all five bugs detected by FullSet were also detected by \cfgset{}, further emphasizing the superiority of initial seed selection.

We also conducted a large-scale fuzzing experiment by applying FullSet and \cfgset{} for each corpus to VECT with a total of over 2,000 hours, which aimed to evaluate the time overhead required by FullSet to detect the same number of unknown bugs as \cfgset{}.
We found that FullSet detecting all the bugs detected by \cfgset{} required over 1,000 hours in total.
This further demonstrates that the initial seed selection method \cfgset{} significantly improves JVM fuzzing performance.

\begin{table*}[t]
\caption{Comparison results of different methods in terms of the number of previously unknown bugs}
\small
\begin{spacing}{1.2}
\centering
\resizebox{0.85\linewidth}{!}{
\begin{tabular}{cccccccccccc}
\toprule
 \textbf{Method}     
 &\textbf{FullSet} 
 &\textbf{\minset{}} 
 & \textbf{\peachset{}} 
 & \textbf{\hotset{}} 
 & \textbf{\tfidfset{}} 
 & \textbf{\astset{}}
 & \textbf{\cfgset{}}
 & \textbf{\codebertset{}}
 & \textbf{\infercodeset{}}
 & \textbf{\plbartset{}}
 & \textbf{\codetset{}}
 \\ \midrule
 
\textbf{\begin{tabular}[c]{@{}c@{}}\#Unknown\\ Bugs\end{tabular}} 
& 5 & 6 & 6 & 5 & 6 & 8 & \cellcolor[HTML]{BBBBBB}11 & 8 & 7 & 5 & 7       
\\ \bottomrule
\end{tabular}
}
\end{spacing}
\vspace{-.4cm}
\label{table:bug_count}
\end{table*}

\begin{figure}
    \centering
    \includegraphics[width=.88\linewidth]{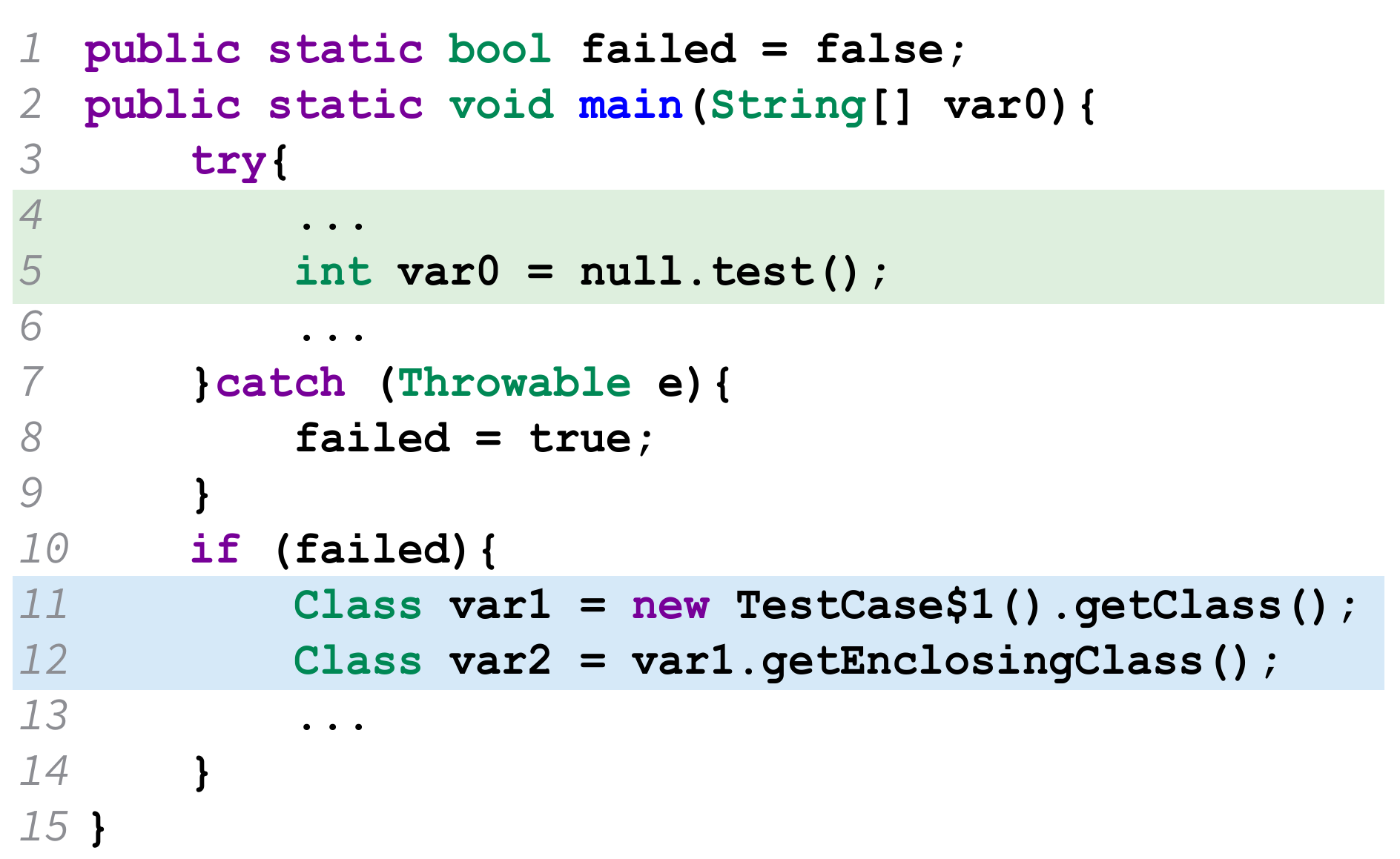}
    \caption[OpenJ9 bug \#19016]{OpenJ9 bug \#19016}
    \label{fig:testcase}
    \vspace{-.4cm}
\end{figure}

In total, \unknownBug{} previously unknown bugs were detected by our initial seed selection methods in our experiments, \confirmBug{} of which have been confirmed and fixed by developers.
Table~\ref{tab:unknown_bug} shows the information about these \confirmBug{} bugs.
Among these, \newCorpusBug{} were bugs that could only be detected by P3, further highlighting the significance of the open-source corpus in enhancing JVM fuzzing techniques.
We then used a previously unknown bug detected by P3 as an example to illustrate the effectiveness of open-source corpus. 
Figure~\ref{fig:testcase} shows a simplified synthesized test program that triggers a runtime check bug\cite{bugCase} in OpenJ9, as detected by P3. 
In this example, lines shaded green (Lines 4,5) represent the ingredient that was inserted, while lines shaded blue (Lines 11,12) represent the root cause of the bug. The original seed program code does not throw any exception, so the {\tt failed}'s value is false, which causes the branch containing the bug not to be executed. VECT inserts an ingredient containing a {\tt NullPointerException} which causes the value of {\tt failed} to be changed to true (Line 8). In the branch, the synthetical program first news an anonymous inner class (i.e., {\tt TestCase\$1}) and tries to get its enclosing class (Line 12). If {\tt TestCase\$1} is a normal inner class, OpenJ9 needs to check whether the inner class and its enclosing class have the same {\tt InnerClass} attribute. However, OpenJ9 incorrectly applies this check to anonymous inner classes and throws an {\tt IncompatibleClassChangeError}. The developers of OpenJ9 have confirmed and fixed this bug. Note that this bug only can be detected by P3, since this bug requires the seed program to new an anonymous inner class and call the {\tt getEnclosingClass} function.

%\footnotemark
% \footnotetext{https://github.com/eclipse-openj9/openj9/issues/19016}

\begin{tcolorbox}[colback=gray!5]
\textbf{Finding V}:
Initial seed selection enhances JVM techniques for detecting more previously unknown bugs given the same testing period, with \confirmBug{} being already confirmed and fixed by developers.
Seven of them can only be found by the open-source corpus.
% Given the same testing period, \cfgset{} can detect more unknown bugs than the full set of initial seeds.
\end{tcolorbox}

% \section{Discussion}

\section{Implications and Future Work}
% \jj{Litter new information is provided in this section compared to our results and observations. More actionable suggestions are needed.}

% \jj{Some example points:
% (1) We recommend the most practical method by considering effectiveness and efficiency: CFG-based method.
% we can thus highlight the importance of designing task-specific methods.
% (2) Different corpus exhibit complementary effectiveness. The existing work always used the same corpus for fuzzing even though some new techniques are proposed. It is important to consider the diversity of corpus during fuzzing. In particular, we may further mix all corpus for further selection and integrate their diverse testing capability for improving the overall effectiveness.
% (3) Boosting the effectiveness of code representation based methods through fine-tuning may be helpful, inspired by the different effectiveness of the three corpus.
% (4) I'm not sure whether we should mention that different selection methods may be complementary in bug detection. Then, we disucss their integration potential.
% (5) I wonder whether smaller scale (20\%) can have better effectiveness than larger scale (50\%) at the beginning period of fuzzing. If so, we can also recommend the scale according to the given time budget. That is, we should show the influence of time budget on bug detection.}

\textbf{Designing a task-specific initial seed selection method is crucial.}
As shown in RQ1 and RQ2, methods widely evaluated in traditional software testing are not suitable for JVM testing due to the characteristics of large-scale and intricate code in JVM.
Inspired by existing work, we propose program-feature-based selection methods and demonstrate their superiority.
These findings suggest a need to design task-specific seed selection methods, as different test subjects often exhibit unique characteristics.
For efficiency and effectiveness, we recommend using the \cfgset{} method, which demonstrated the most practical in our JVM testing experiments.
% Because different subjects have different characteristics, designing task-specific initial seed selection methods is crucial. 

\textbf{Selecting an optimal corpus budget is significant.}
The results shown in Table~\ref{tab:budgets} indicate that different initial seed selection methods tend to achieve the best performance at varying corpus budgets.
Specifically, we recommend using \cfgset{} with a 50\% budget for initial seed selection, which has been found to optimize performance and resource utilization in JVM fuzzing.
Future work could focus on refining and implementing an adaptive budget selection algorithm for \cfgset{}, in order to determine the minimum corpus budget required for the best performance.

\textbf{Exploring open-source corpus for improving JVM fuzzing is beneficial.}
Existing work often used the same corpus as older works for fuzzing, even when proposing new techniques, thus neglecting the significance of the diverse corpus.
Our RQ2 results confirmed that different corpus exhibit complementary effectiveness by detecting new behaviors and bugs.
Significantly, the findings reveal that the initial seed selection can improve the quality of the open-source corpus.
Hence, future work could further mix all corpus for further selection and integrate their diverse testing capability to enhance overall effectiveness.
% Moreover, initial seed selection not only improves the quality of open-source corpus but also allows for further blending of all corpus for additional selection, integrating their diverse testing capabilities to enhance overall effectiveness.

\textbf{Fine-tuning the pre-trained model for initial seed selection shows promise.}
As shown in table~\ref{tab:unique_inconsistency}, pre-trained-model-based \feabase{} achieves substantial performance enhancements on the open-source corpus compared to the benchmark corpus. 
This is because the open-source corpus data is collected from GitHub, which enhances the learning of code semantics. 
To further improve effectiveness, future work will involve fine-tuning the pre-trained code representation models across various programming languages and compilers to better fit the downstream task of initial seed selection.

\section{Threats to Validity}
\textit{External} threats to validity primarily lie in the corpus and fuzzers used in our study. 
Firstly, we eliminated seed programs that can directly identify inconsistencies to prevent interference with fuzzing results. In the future, we plan to incorporate additional corpus to further mitigate the threat. 
% Secondly, there are false positives in the result of the fuzzer. To reduce the threat of false positives, more filtering rules are designed and applied in VECT and JavaTailor to obtain more accurate results.
Secondly, although these initial seed selection methods can be generalized to any JVM fuzzer that takes seeds as input, we only evaluated the impact of initial seed selection on JavaTailor and VECT.
We selected them as the representatives due to their state-of-the-art effectiveness~\cite{vect,zang2024} and general testing purposes (e.g., JITFuzz~\cite{jitfuzz} targets the JIT component only).

\textit{Construct} threats to validity mainly stem from the duration of each fuzzing process and its randomness.
Despite our best efforts to assess the impact of initial seed selection and open-source corpus on JVM fuzzing effectiveness, the fuzzing time we have set is considerably shorter than that in industrial fuzzing. Due to resource constraints, we are unable to extend the fuzzing time for each iteration.
Besides, the execution of the fuzzer and breaking ties involve randomness. To reduce the impact of this randomness, we performed each experiment five times using different random seeds.

\textit{Internal} threat to validity mostly lies in the implementations of each technique. 
To mitigate this threat, we relied on the available APIs to extract AST and CFG and utilized the pre-trained code representation models with their default settings.
% \feabase{} needs to manually set the selection budget, in the future work we will design an algorithm to solve this threat. \Wang{TODO}

% \input{7_related_work}
\section{Conclusion}
This work designed a total of 10 initial seed selection methods (i.e., coverage-based, prefuzz-based, and program-feature-based), in order to enhance the JVM fuzzing effectiveness.
We conducted an empirical study on three JVM implementations with JavaTailor and VECT to comprehensively evaluate the performance of initial seed selection methods.
The results highlight that the subset selected by initial seed selection outperforms the entire set of initial seeds. 
In particular, the method utilizing control flow graphs performs the best.
Furthermore, the study emphasizes the benefits of incorporating programs in the wild, demonstrating the complementary effectiveness with the existing benchmark corpus.
In addition, given the same testing period, initial seed selection can enhance fuzzing techniques by detecting more previously unknown bugs, with \confirmBug{} bugs having been confirmed or fixed by developers.
Our work also opens up several promising future directions including determining the minimum corpus budget, fine-tuning pre-trained code representation models for better fitting the downstream task, and mixing diverse corpus for selection.

\noindent
\textbf{Data Availability.} The replication package that supports the findings of this study is available publicly~\cite{homepage}.

% In order to improve the efficiency of existing initial seed selection methods, and evaluate its impact on JVM test performance. We adapt two kinds of existing methods \covbase{} and \fuzzbase{}, and design a black box method \feabase{} inspired by TCP technology. Its key insight is that similar programs are likely to trigger similar defects, so it prioritizes selecting seed programs that contain special features. To make the evaluation more complete and JVM testing more effective, we collected new corpus from the open source community to further evaluate the impact of open source corpus on JVM fuzzing performance. We conducted extensive study on three popular JVM implementations (HotSpot, OpenJ9, and Bisheng JDK) using two JVM fuzzers (JavaTailor and VECT). The results demonstrate that CFG feature-based \feabase{} can detect the most inconsistencies. In particular, we identified a total of \unknownBug{} previously unknown bugs, out of which \confirmBug{} bugs have been confirmed or fixed by developers, and \newCorpusBug{} bugs were only detected by the collected open-source corpus.
% \Wang{TODO}

\section*{Acknowledgment}
We thank all the ICSE anonymous reviewers for their valuable comments. We also thank all the JVM developers for analyzing and replying to our reported bugs. 
The work has been supported by the National Natural Science Foundation of China Grant Nos. 62322208, 12411530122, 62232001, CCF Young Elite Scientists Sponsorship Program (by CAST), and Huawei Fund. 

% \begin{acks}
% temp acknowledgments
% \end{acks}

% \newpage
\balance
\bibliographystyle{IEEEtran}
\bibliography{refer}

\end{document}